\documentclass[a4paper,nobibnotes,nofootinbib]{revtex4}

\usepackage{amssymb}
\usepackage{amsmath}
\usepackage{epsfig}
\newcommand{\be}{\begin{equation}}
\newcommand{\ee}{\end{equation}}
\newcommand{\bea}{\begin{eqnarray}}
\newcommand{\eea}{\end{eqnarray}}

\begin{document}
\title{Search for exclusive events using the dijet mass fraction at the Tevatron}
\author{O. Kepka}\email{kepkao@fzu.cz}
\affiliation{DAPNIA/Service de physique des particules, CEA/Saclay, 91191
Gif-sur-Yvette cedex, France}
\affiliation{IPNP, Faculty of Mathematics and Physics,
Charles University, Prague} 
\affiliation{Center for Particle Physics, Institute of Physics, Academy of Science, Prague} 
\author{C. Royon}\email{royon@hep.saclay.cea.fr}
\affiliation{DAPNIA/Service de physique des particules, CEA/Saclay, 91191 
Gif-sur-Yvette cedex, France}

\begin{abstract}

In this paper, we discuss the observation of exclusive events using the dijet
mass fraction as measured by the CDF collaboration at the Tevatron. We compare
the data to pomeron exchange inspired models as well as Soft color interaction
ones. We also provide the prediction on dijet mass fraction at the LHC using
both exclusive and inclusive diffractive events. 
\end{abstract}
\maketitle

\newlength{\picwidth}
\setlength{\picwidth}{8.9cm}
\renewcommand{\arraystretch}{1.5}

\section{Introduction}

Double pomeron exchange (DPE) processes are expected to extend 
the physics program at the LHC not only due to the possible Higgs 
boson detection but also because of the possibility to study broader range 
of QCD physics and diffraction \cite{sensitivityHiggs, Royon:2003ng, higgsInDPE, susy, 
HiggsGamma, Khoze:2007hx,Khoze:2006iw, bpr}. The processes are theoreticaly characterized by large 
rapidity gap regions devoid of particles between centrally produced heavy object and
the scattered hadrons which leave the interaction intact. This is attributed to the 
exchange of a colorless object, the  pomeron (or the reggeon). In the LHC environment, however, the 
rapidity gap signature will not appear because of the high number of multiple interactions 
occuring at the same time, and the diffractive events will be identified tagging the escaping
protons in the beam pipe. 

One generally considers two classes of DPE processes, namely, \textit{exclusive} DPE events 
if the central object is produced alone carrying away the total 
available diffractive energy and \textit{inclusive} events when the 
total energy is used to produce the central object and in addition the
pomeron remnants. Exclusive events allow a precise reconstruction of the
mass and kinematical properties  of the central object
using the central detector or even more precisely  using very forward detectors
installed far downstream from the interaction point. The most appealing exclusive process to be 
studied at the LHC is the Higgs boson production but since it cannot be observed at the Tevatron
due to the low production cross section, one should
find other ways to look for exclusive events at the Tevatron, for example 
in dijet, diphoton channels.
It is needed to mention that until recently, there was not a decisive measurement that would provide enough evidence for the existence of exclusive 
production.

Although exclusive production yields kinematically well constrained final state objects, their experimental detection
is non-trivial due to the overlap with the \textit{inclusive} DPE events. 
In those events, the colliding pomerons are
usualy viewed as an object with partonic sub-structure. A parton emitted from the pomeron takes part in the 
hard interaction and pomeron remnants accompanying the central object are distributed uniformly in rapidity. Exclusive events usually appear as a small deviation from the inclusive model predictions which need to be
studied precisely
before accepting a new kind of production. In particular, the structure of the pomeron
as obtained from HERA is not precisely known at high momentum fraction, and specifically, the gluon in the pomeron is not 
well constrained. It is not clear if such uncertainty could not lead to mis-identifying
observed processes as exclusive. This would for instance preclude the  spin analysis of the produced 
object. 
\par
In this paper, we aim to investigate the observation of exclusive production 
at the Tevatron. Indeed, we use the dijet mass fraction distribution measured 
by the CDF collaboration and show
that even taking into account uncertainties associated with the pomeron structure, one is unable to give
a satisfactory description of the data without the existence of exclusive events. 
We also include other approach to explain diffraction in our study, the so called Soft color interaction 
model 
(the properties of all the models are discussed later). 
As an outlook, 
we apply current models for the DPE production for LHC energies and demonstrate the possible appereance 
of exclusive events through the dijet mass fraction. 
\par
The paper is organized as follows: in the second section we give a brief description of the inclusive, exclusive, and
Soft color interaction models.
The third section 
discusses how well the various models can explain the preliminary
Tevatron dijet mass fraction data and the constraints 
implied by data on the current models. 
In the fourth part, we foreshadow an application of the dijet mass fraction 
distribution as a tool to observe exclusive events at LHC energies. Finally, we discuss issues concerning the dijet mass fraction reconstruction and fast detector simulation in the Appendix.

\section{Theoretical Models}

Inclusive and exclusive DPE models used in this paper are implemented in the Monte Carlo program {DPEMC} \cite{dpemc}. 
The Soft color interaction model is embeded in the PYTHIA program \cite{SCIMC}. 
The survey of the different models follows.

\subsection{Inclusive Models}

The first inclusive model to be mentioned  is the so called ``Factorized model". It is 
an Ingelman-Schlein type of model \cite{IS}
describing the diffractive double pomeron process as a scattering of two pomerons emitted from the proton, assuming a 
factorization of the cross section into a regge flux convoluted with the pomeron structure
functions. For $ep$ single diffraction, it is necessary to introduce secondary reggeon trajectory to describe the 
observed single diffractive non-factorable cross section. In the case of the 
Tevatron, the pomeron trajectory alone is sufficient to describe present data
and the cross section is factorable as it was advocated in
\cite{factorization}. Factorization breaking between HERA and Tevatron 
comes only through the survival probability factor, denoting the probability that there is no additional 
soft interaction which would destroy the diffractively scattered protons. 
In other word, the probability to destroy the rapidity gap does not depend
on the hard interaction.
At Tevatron energies, the factor
was measured to be approximately 0.1, and calculation suggested the value of 0.03 for the LHC.
Pomeron structure functions, reggeon and pomeron fluxes are determined from 
the DIS $ep$ collisions fitting the 
diffractive structure function $F_2^{D}$ at HERA. For one of the most 
recent published diffractive structure function analysis see e.g \cite{pdfs}.

\par On the other hand, the Bialas-Landshoff (BL) inclusive model \cite{bpr}, is a purely non-perturbative
calculation utilising only the shape of the pomeron structure function and leaving the overall 
normalization to be determined from the experiment; one can for example confront the prediction of 
DPE cross section with the observed rate at the Tevatron \cite{factorization} and 
obtain the missing normalization factor \footnote{One more remark is in order.
In the BL inclusive model, the partonic content of the pomeron is expressed in terms of the distribution functions
as $f_{i/\mathbb{P}}(\beta_i)\equiv \beta_i G_{i/\mathbb{P}}(\beta_i)$, 
where the $G_{i/\mathbb{P}}(\beta_i)$ are the true parton densities
as measured by the HERA collaboration, and $\beta_i$ denotes the momentum fraction of the parton $i$ in the pomeron.
The integral of $f_{i/\mathbb{P}}(\beta_i)$ is normalized to 1, so that in the 
limit $f_{i/\mathbb{P}}(\beta_i)\rightarrow \delta(\beta_i)$ the exclusive cross 
section is recovered \cite{dpemc}.}.
\par
Both models use the pomeron structure measured at HERA which is gluon
dominated. In this paper, we use the results of the QCD fits to the most
recent Pomeron structure function data measured by the H1 collaboration
\cite{pdfs}. The new gluon density in the Pomeron is found to be slightly
smaller than the previous ones, and it is interesting to see the effect
of the new PDFs with respect to the Tevatron measurements.
However, the gluon density at high $\beta$, where $\beta$ denotes the 
fraction of the particular parton in the pomeron, 
is not well constrained from the QCD fits performed at HERA. 
To study this uncertainty, we multiply the gluon distribution by the
factor $(1 - \beta)^{\nu}$ as shown in Fig. 1. QCD fits to the H1 data lead to 
the uncertainty on the $\nu$ parameter $\nu=0.0\pm0.5$ 
\cite{pdfs}. We will see in the following how this parameter influences the
results on dijet mass fraction as measured at the Tevatron.

\subsection{Exclusive Models}

Bialas-Landsoff exclusive model \cite{BLexc} is based on an exchange of two 
``non-perturbative" gluons between a pair of colliding hadrons which 
connect to the hard subprocess. Reggeization is employed in 
order to recover the pomeron parameters which successfully
described soft diffractive phenomena, e.g. total cross section at low energies. 
A calculation of $q\bar{q}$ and $gg$ production and more details can be found 
in \cite{BLexc} and \cite{BLgg}, respectively. 

On the contrary, the Khoze, Martin, Ryskin (KMR) \cite{kmr} model is purely a 
perturbative approach. The interaction is obtained by an exchange of two gluons directly coupled to the colliding hadrons (no pomeron picture is introduced). While one gluon takes part in the creation of the central object, the other serves
to screen the color flow across the rapidity gap. 
If the outgoing protons remain intact and scatter at small angles, 
the exchanged di-gluon system, in both models, must obey the selection rules $J_{Z}=0$, C-even, P-even. Such constrains are also applied to the hard subprocesses for the production of the central object. 
\par
The two models show a completely different $p_T$ dependence of 
the DPE cross section. The energy dependence of the BL model is found
to be weaker since the Pomeron is assumed to be soft whereas it is not the case
for the KMR model.

\subsection{Soft Color Interaction Model}
The Soft color interaction model (SCI) \cite{sci,SCIMC} assumes that diffraction
is not due to a colorless exchange at the hard vertex but
rather to a string rearrangement in the final state during hadronisation. 
This model gives a probability (to be determined by the experiment) that
there is no string connection, and so no color exchange, between the partons
in the proton and the scattered quark produced during the hard interaction.
Since the model does not imply the existence of a pomeron, there is no need of a
concept like survival probability and a correct normalisation is found 
between single diffraction 
Tevatron and HERA data without any new parameter, which is one of the
big successes of this model.

\begin{figure}
\includegraphics[width=1.5\picwidth]{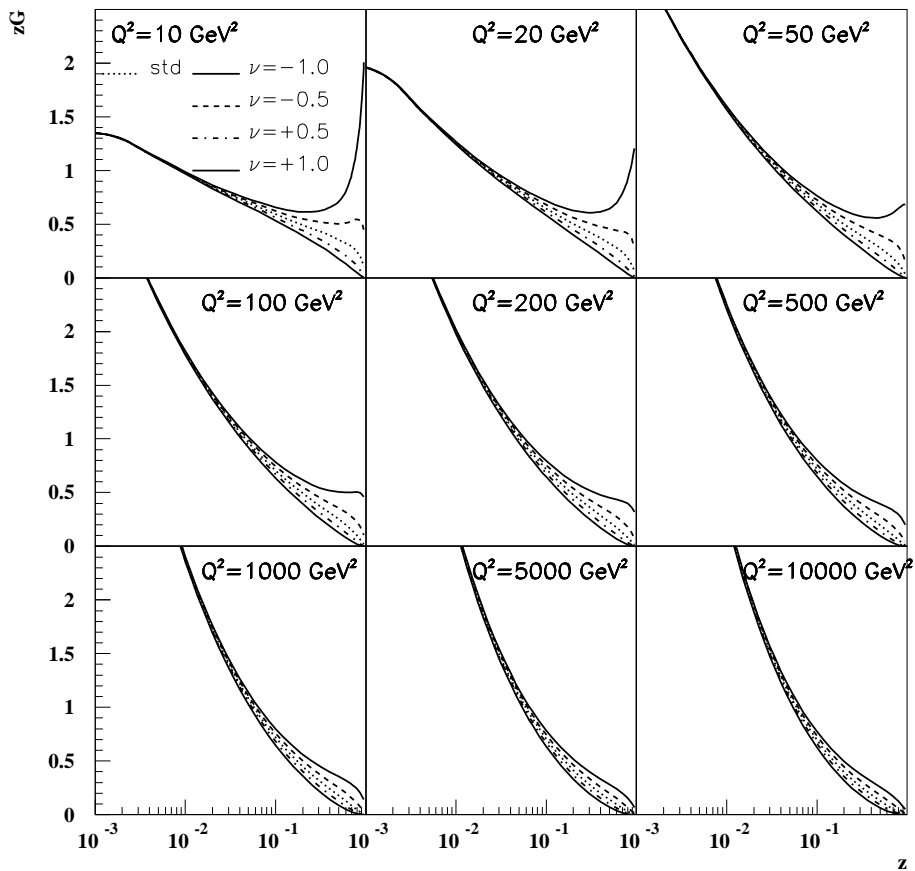}
\caption{Uncertainty of the gluon density at high $\beta$ (here $\beta\equiv z$).
The gluon density is multiplied by the factor $(1-\beta)^{\nu}$ where $\nu$=-1.,
-0.5, 0.5, 1. The default value $\nu =0$ is the gluon density in the pomeron
determined directly by a fit to the H1 $F_2^D$ data with an uncertainty
of about 0.5.}
\label{gluon}
\end{figure}
\section{Dijet mass fraction at the Tevatron}
\label{sect:dmf}

Dijet mass fraction (DMF) turns out to be a very appropriate observable for identifying the exclusive 
production. It is defined as a ratio $R_{JJ}=M_{JJ}/M_{X}$ of the dijet system 
invariant mass $M_{JJ}$ to the total mass of the final state system $M_X$ (excluding the intact beam (anti)protons).
If the jet algorithm has such properties that the outside-cone effects are small, 
the presence of an exclusive production would manifest itself as an excess of the 
events towards $R_{JJ}\sim1$; for exclusive events, the dijet mass is essentially equal
to the mass of the central system because no pomeron remnant is present. 
The advantage of DMF is that one can focus on the shape of the distribution; the observation 
of exclusive events does not rely on the overall normalization which might be strongly dependent on
the detector simulation and acceptance of the roman pot detector.
\par
In the following analysis, we closely follow the measurement performed by the CDF Collaboration. 
One can find more information about the measurement and the detector setup in a note discussing 
preliminary results \cite{cdfnote}. In this paragraph,
we will mention only the different cuts which are relevant for our analysis. 
To simulate the CDF detector, we use a fast simulation interface \cite{fastSimul}, which performs a smearing of the
deposited cell energy above a $0.5\,\mathrm{GeV}$ threshold and reconstructs jets using a cone algorithm. 
Properties of the event such as the rapidity gap size were evaluated at
the generator particle level.

CDF uses a roman pot detector to tag the antiprotons on one side (corresponding  to $\eta_{\bar{p}} < 0$). For the DMF reconstruction, 
we require the antiprotons to have the longitudinal momentum loss in the range $0.01<\xi_{\bar{p}}<0.12$ and we
apply the roman pot acceptance obtained from the CDF Collaboration (the real acceptance is greater than 0.5 for 
$0.035< \xi_{\bar{p}}<0.095$).
On the proton side, where no such device is present, 
a rapidity gap of the size $3.6<\eta_{gap}<5.9$ is required. In the analysis, 
further cuts are applied: two leading jets with a transverse
momentum above the threshold $p^{jet1,jet2}>10\,\mathrm{GeV}$ or $p^{jet1,jet2}_T>25\,\mathrm{GeV}$ in
the central region $|\eta^{jet1,jet2}|<2.5$, a third jet veto cut  
($p_T^{jet3}<5\,\mathrm{GeV}$) 
as well as an additional gap on the antiproton side of the size $-5.9<\eta_{gap}<-3.6$. For the sake of brevity, the threshold for the
transverse momentum of the two leading jets will be in the following  denoted as $p_T^{min}$, if needed.

The dijet mass  is computed using the 
jet momenta for all events passing the above mentioned cuts. In order to follow as much as possible the method used by the
CDF collaboration, the mass
of the diffractive system $M_X$ is calculated from the longitudinal antiproton momentum loss 
$\xi_{\bar{p}}$ within the roman pot acceptance, and the longitudinal momentum loss of the proton
$\xi_{p}^{part}$ is determined from the particles in the central detector ($-4<
\sim \eta_{part} < \sim 4$), such that:
\begin{eqnarray}
M_X&=&\sqrt{s\xi_{\bar{p}}\xi_p^{part}},\\
\xi^{part}_p&=&\frac{1}{\sqrt{s}}\sum_{particles} p_T \exp^{\eta},\label{eq:xipart}
\end{eqnarray}
summing over the particles with energies higher than $0.5\,\mathrm{GeV}$ in the final state at generator level. To
reconstruct the diffractive mass, $\xi_p^{part}$ was multiplied by a factor 
$1.1$, obtained by fitting the correlation plot between
the momentum loss of the proton at generator level $\xi_p$ 
and $\xi_p^{part}$ at particle level with a straight line. 
\par The DMF reconstruction is deeply dependent on the accuracy of the detector simulation.
Since we are unable to employ the complete simulation in our analysis, we discuss possible effects
due to the various definitions of DMF on the generator and the particle level in the Appendix. 

\begin{figure}[h]
\includegraphics[width=\picwidth]{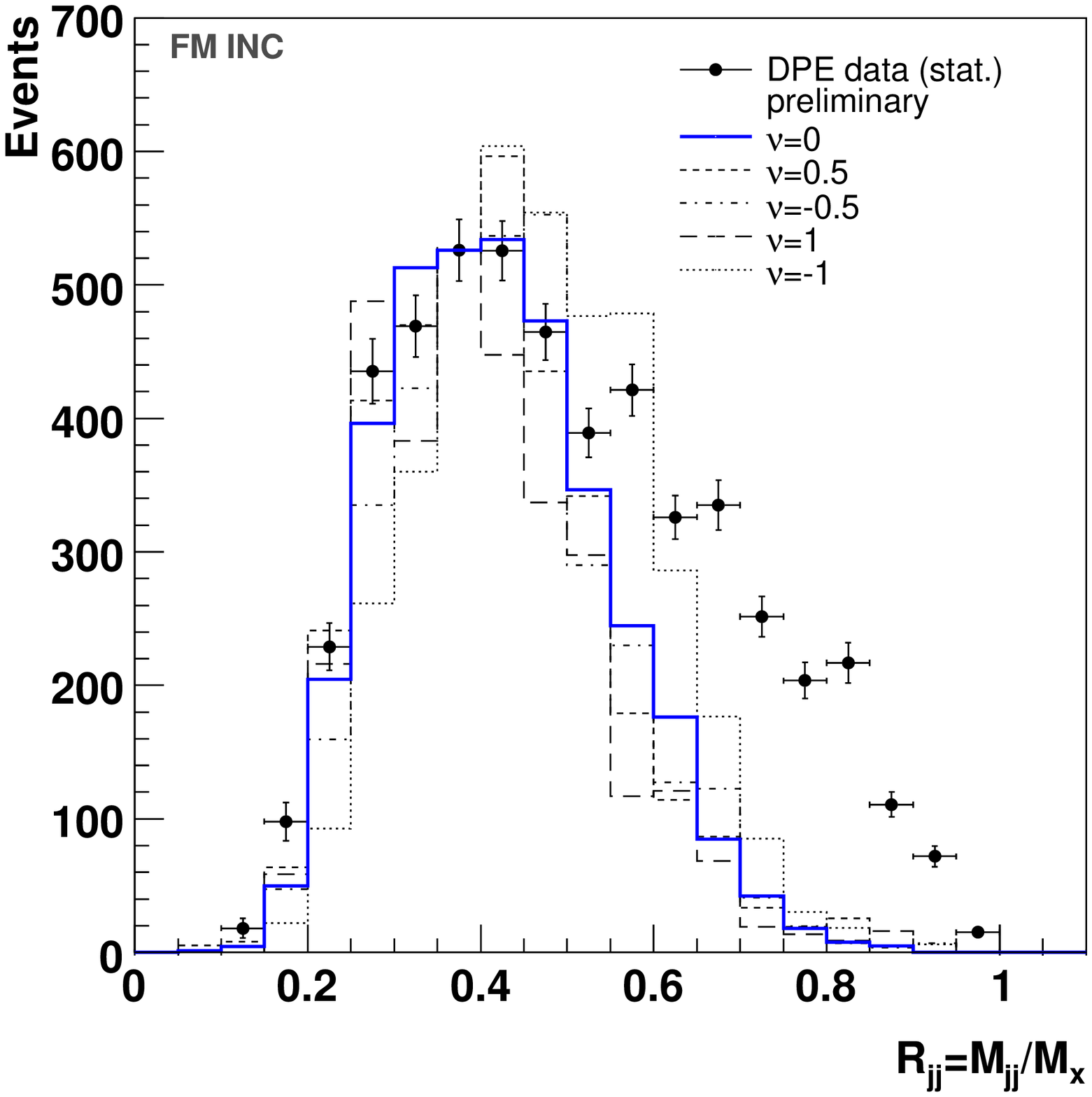}
\includegraphics[width=\picwidth]{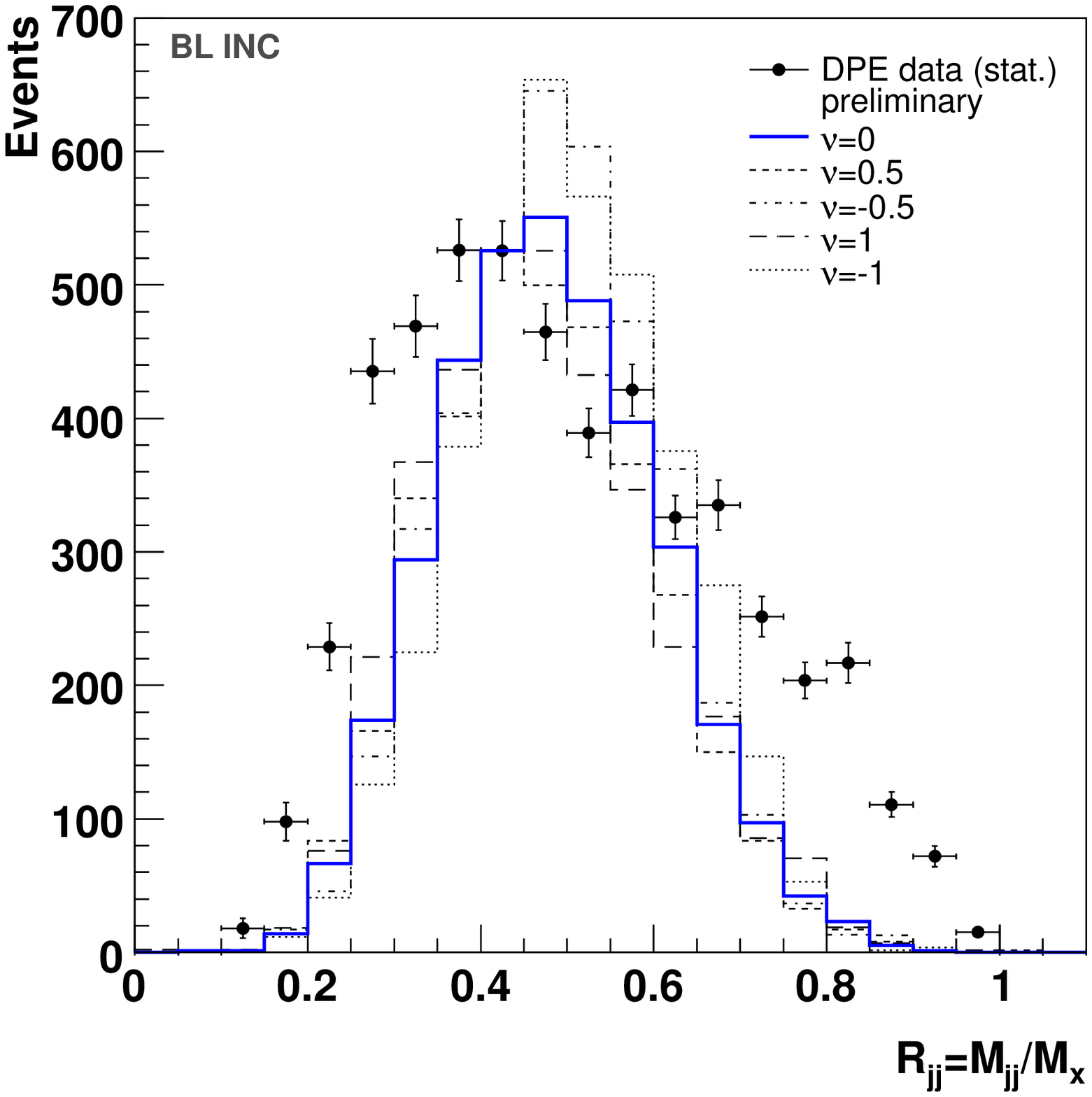}
\caption{Dijet mass fraction for jets $p_T>10\,\mathrm{GeV}$. FM (left) and 
BL (right) models, inclusive contribution. The  
uncertainty of the gluon density at high $\beta$ is obtained by multiplying the 
gluon distribution by $(1-\beta)^{\nu}$ for different values
of $\nu$ (non-solid lines).}
\label{FigInc10}
\end{figure}

\begin{figure}[h]
\includegraphics[width=\picwidth]{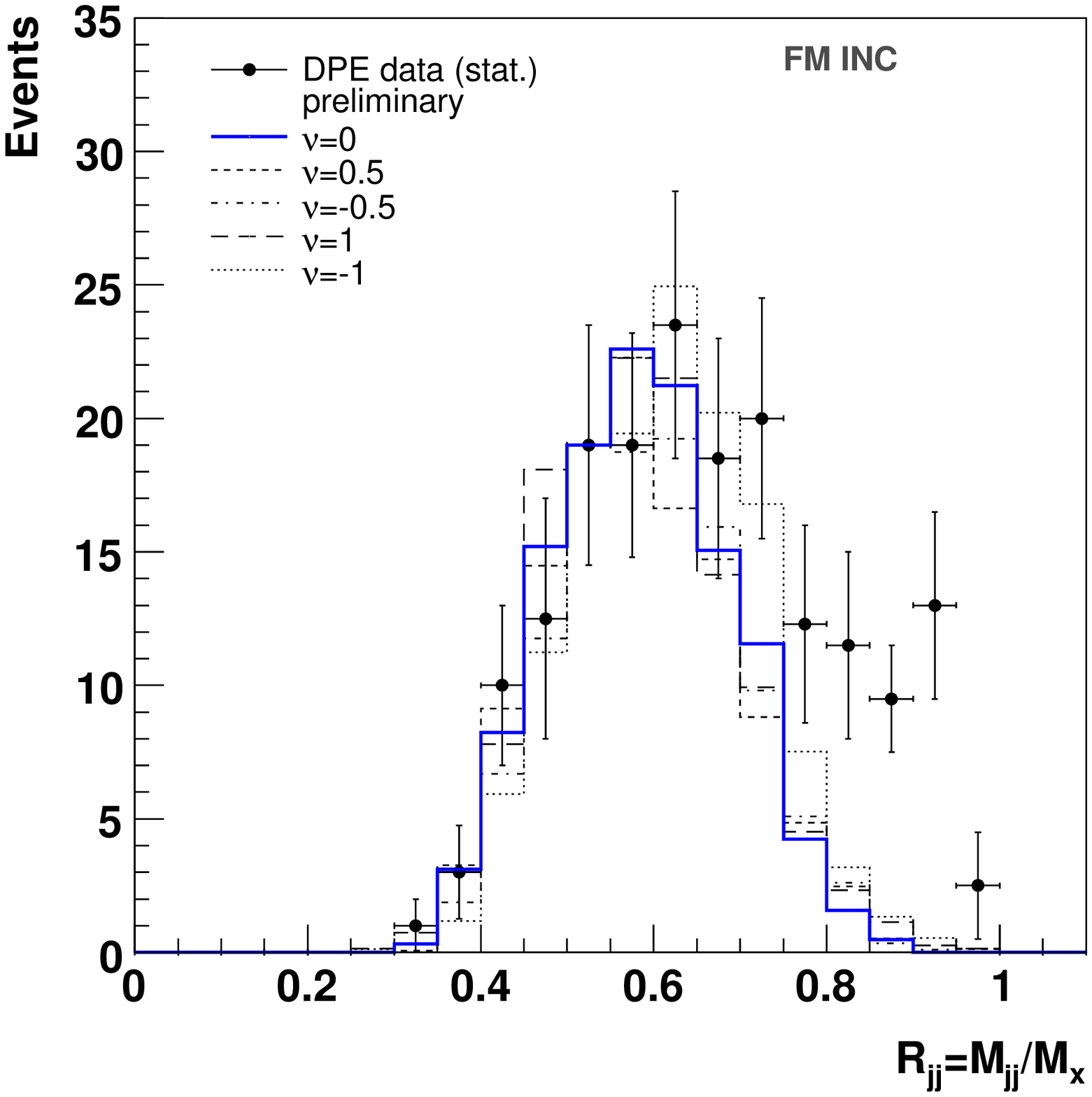}
\includegraphics[width=\picwidth]{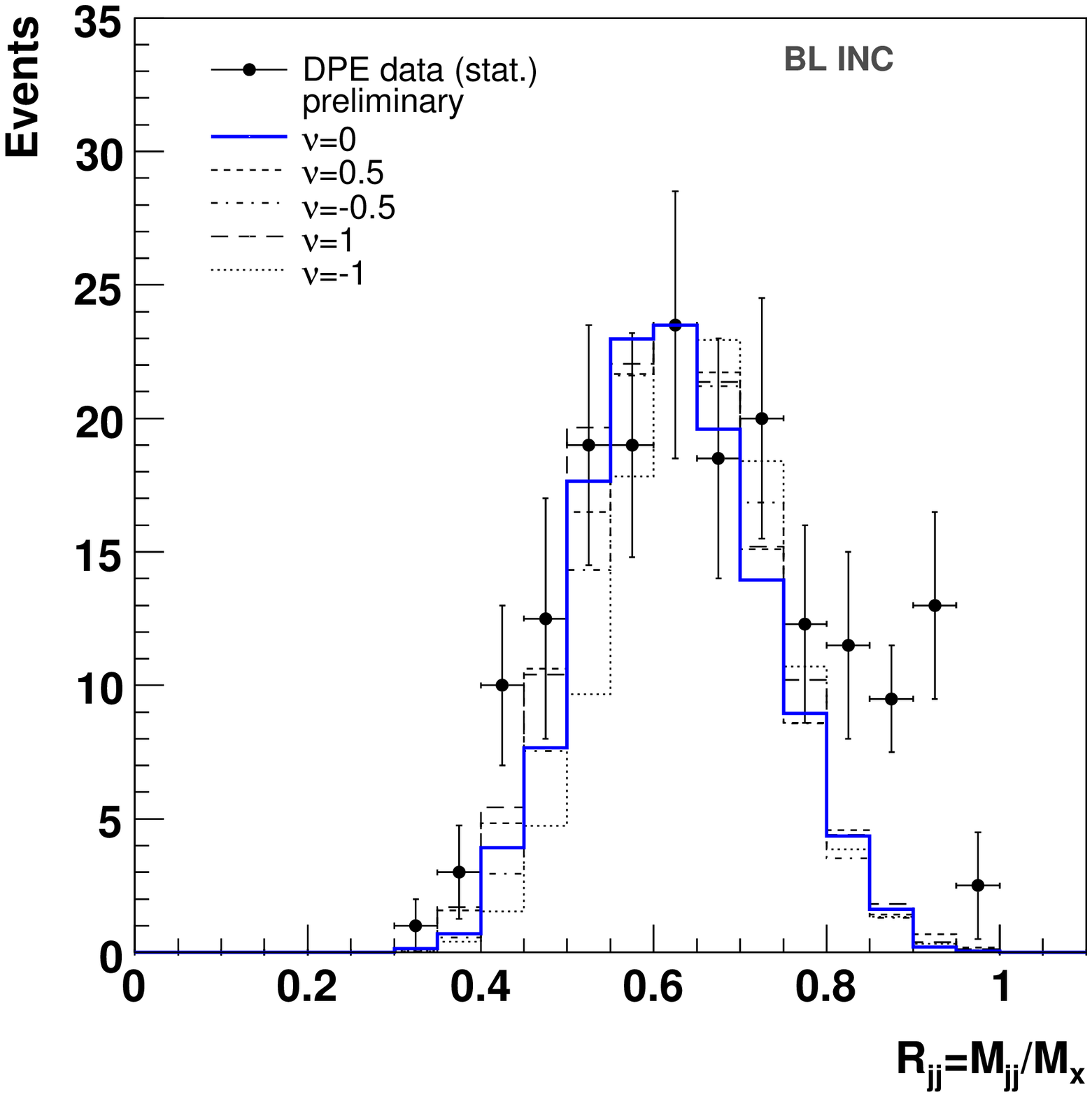}
\caption{Dijet mass fraction for jets $p_T>25\,\mathrm{GeV}$.  FM (left) and 
BL (right) models, inclusive contribution. The  
uncertainty of the gluon density at high $\beta$ is obtained by multiplying the 
gluon distribution by $(1-\beta)^{\nu}$  for different values
of $\nu$ (non-solid lines).}
\label{FigInc25}
\end{figure}

\subsection{Inclusive model prediction}

We present first the dijet mass fraction calculated with FM and BL inclusive models.
As stated in a previous section, we want to explore the impact of the high $\beta$ gluon uncertainty in the pomeron.
To do this, we multiply the gluon density by a factor $(1-\beta)^{\nu}$, for diverse values of $\nu=-1,-0.5,0,0.5,1$. The impact of
the parameter is shown in Fig.~\ref{FigInc10} and Fig.~\ref{FigInc25} for jets with $p_T>10\,\mathrm{GeV}$ and $p_T>25\,\mathrm{GeV}$, respectively. The computed distributions were normalized in shape, since there was no luminosity 
determination, implying no cross section estimation, in the CDF measurement. The interesting possible exclusive region at high $R_{JJ}$ is 
enhanced for $\nu=-1$, however, not in such extent that would lead to a 
fair description of the observed distributions. As a consequence, 
the tail of the measured dijet mass fraction at high $R_{JJ}$
cannot be explained by enhancing the gluon distribution
at high $\beta$, and an another contribution such as exclusive events is
required.
\par A particular property seems to disfavour the BL inclusive model at the Tevatron. Indeed, 
the dijet mass fraction is dumped at low values of $R_{JJ}$, especially for jets $p_T>10\,\mathrm{GeV}$.
Since the cross section is obtained as a convolution of the hard matrix element and the distribution functions, the dumping effect
is a direct consequence of the use of a multiplicative factor $\beta$ 
in the parton density functions in the pomeron (see footnote 1). We will come
back on this point when we discuss the possibility of a revised version of
the BL inclusive model in the following.
\par As we have seen, inclusive models are not sufficient to describe well the measured
CDF distributions. Thus, it opens an area to introduce different types
of proceses/models which give a significant contribution at high $R_{JJ}$.

\begin{figure}[h]
\includegraphics[width=\picwidth]{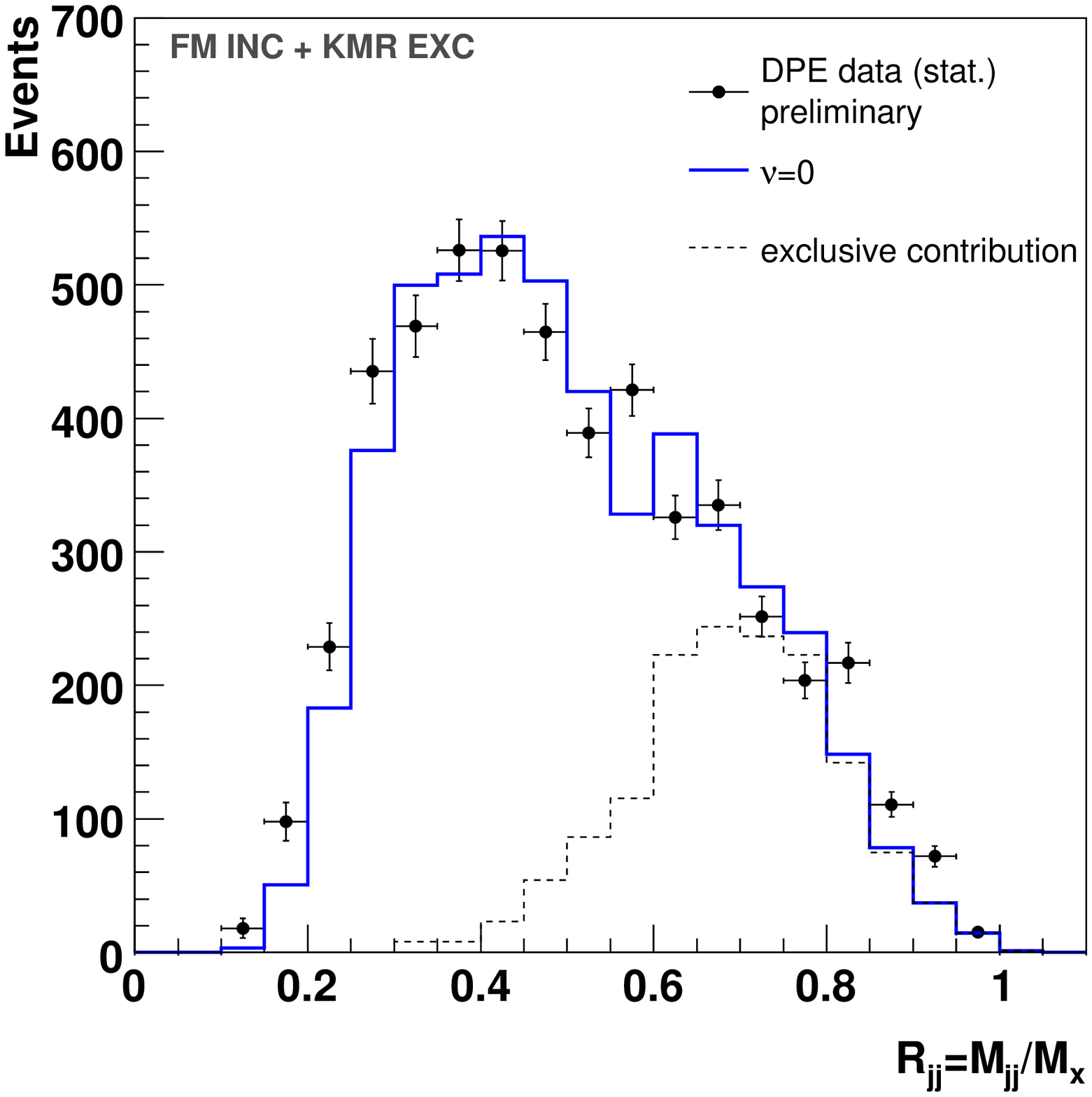}
\includegraphics[width=\picwidth]{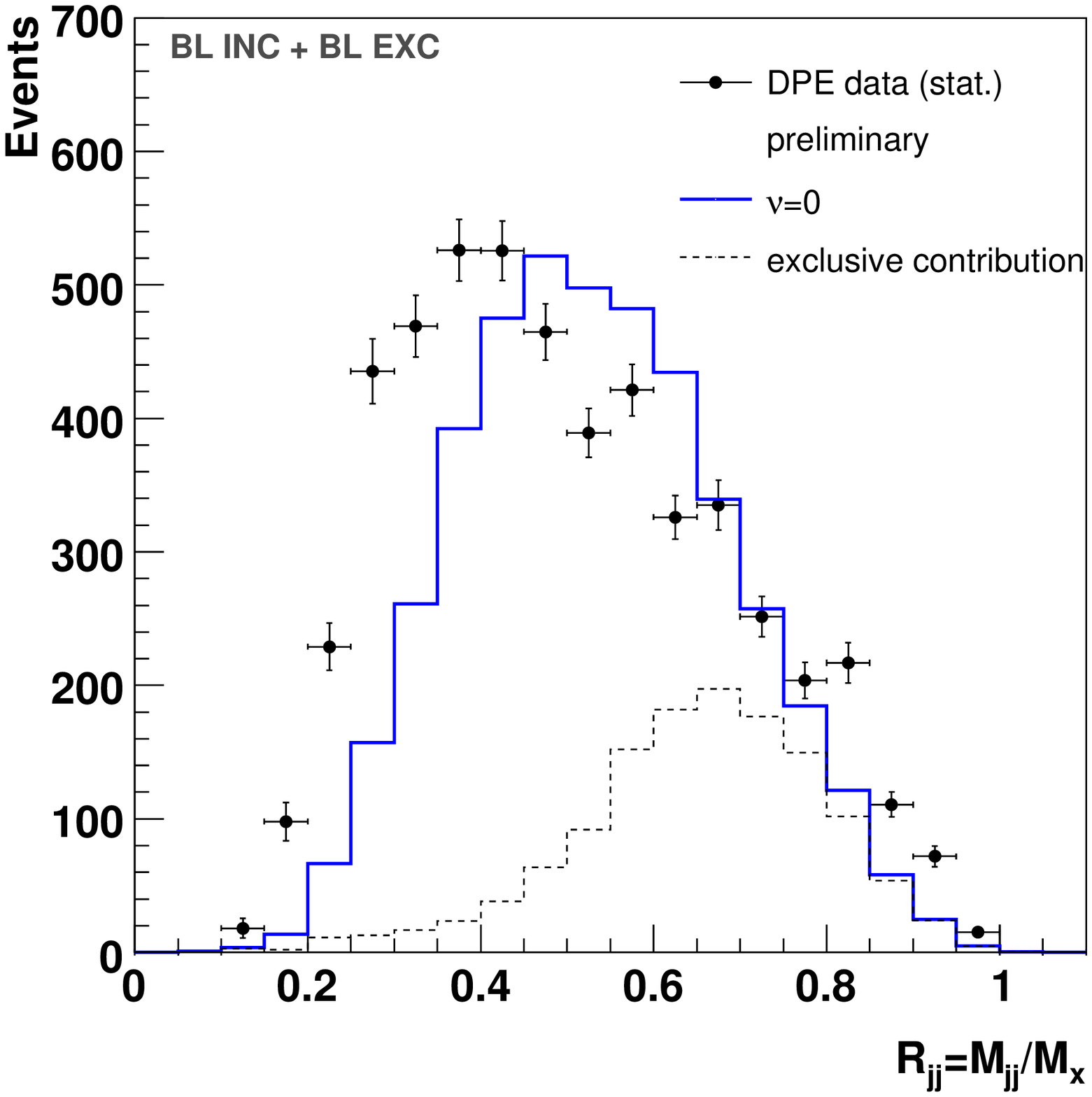}
\includegraphics[width=\picwidth]{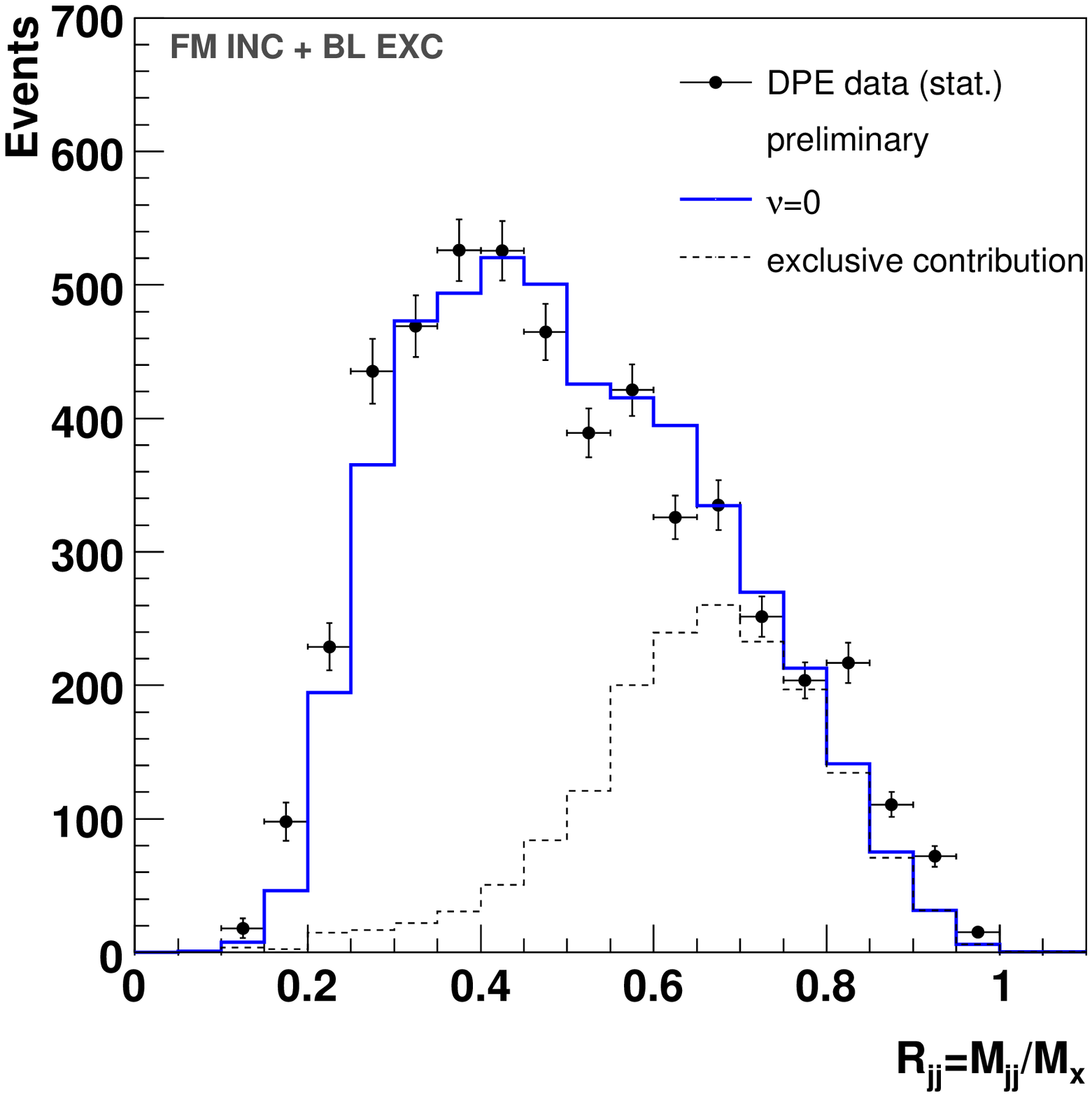}
\caption{Dijet mass fraction for jets $p_T>10\,\mathrm{GeV}$. FM + KMR 
(left), BL + BL (right), FM + BL (bottom) models. We notice that the exclusive
contribution allows to describe the tails at high $R_{JJ}$.}
\label{FigAll10}
\end{figure}

\begin{figure}[h]
\includegraphics[width=\picwidth]{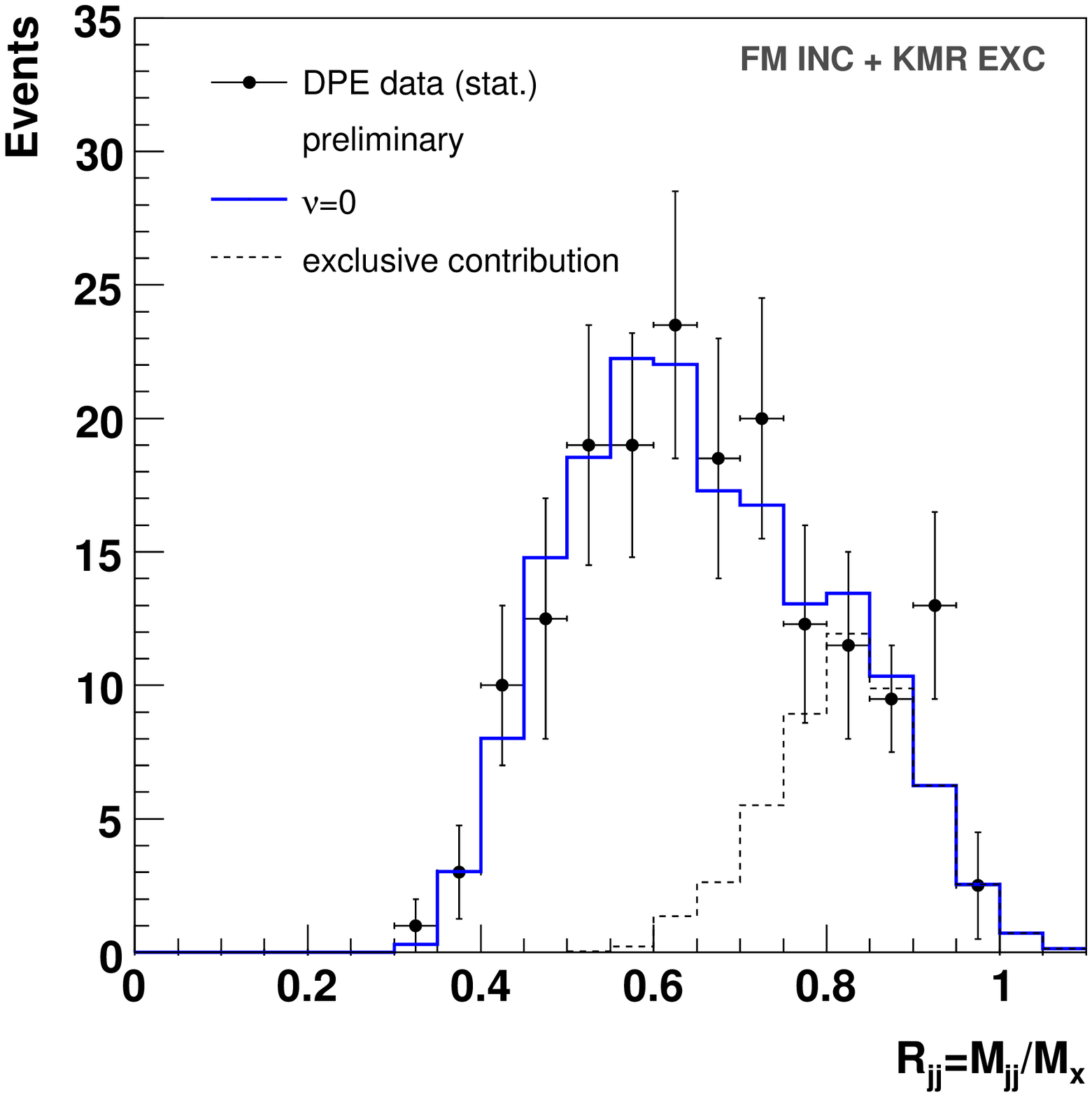}
\includegraphics[width=\picwidth]{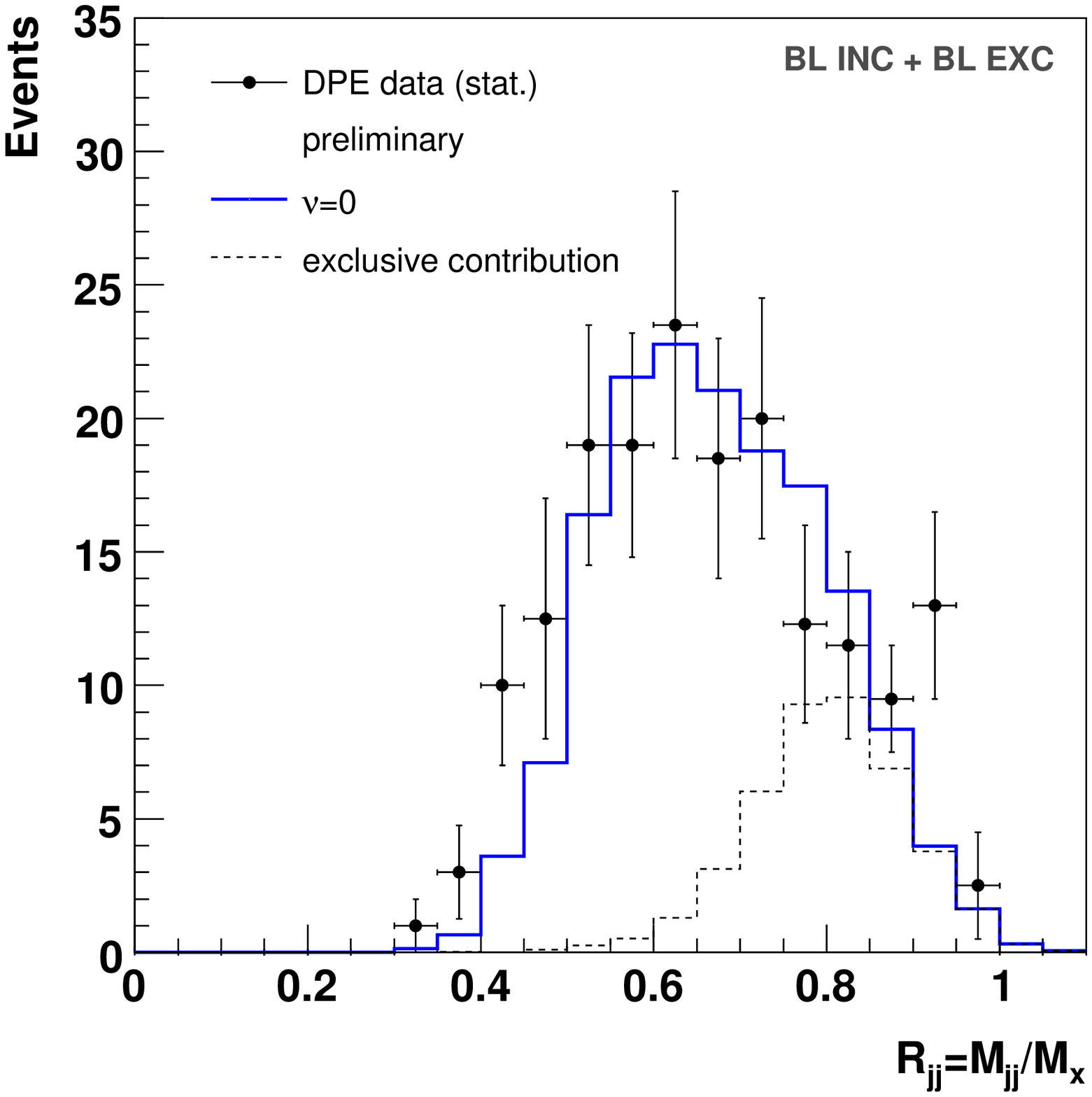}
\includegraphics[width=\picwidth]{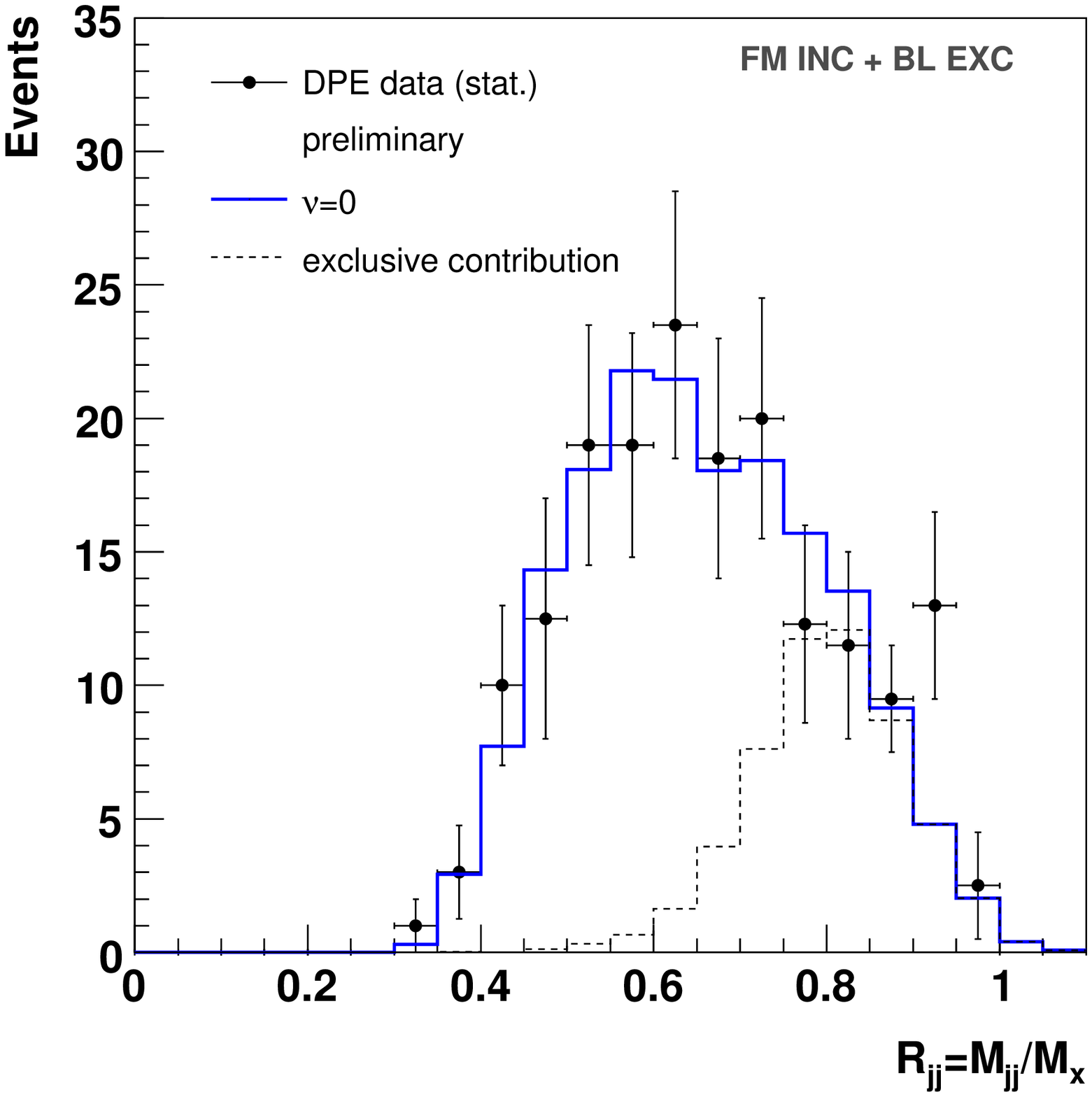}
\caption{Dijet mass fraction for jets $p_T>25\,\mathrm{GeV}$. FM + KMR (left), 
BL + BL (right), FM + BL (bottom) models. We note that the exclusive
contribution allows to describe the tails at high $R_{JJ}$.}
\label{FigAll25}
\end{figure}

\subsection{Exclusive models predictions}

In this section, we will study the enhancement of the dijet mass distribution 
using exclusive DPE processes, with the aim to describe the CDF dijet mass fraction data. We examine 
three possibilities of the interplay of inclusive plus exclusive contributions, specifically:
\begin{enumerate}
\item FM + KMR
\item FM + BL exclusive
\item BL inclusive + BL exclusive
\end{enumerate}
The full contribution is obtained by fitting the inclusive and exclusive distribution to the CDF data, leaving the overall normalization $N$ and the
relative normalization between the two contributions $r^{\mathrm{EXC/INC}}$ free. More
precisely, the DMF distribution is obtained with the fit as $N(\sigma^{\mathrm{INC}}(R_{JJ}) + r^{\mathrm{EXC/INC}}\sigma^{\mathrm{EXC}}(R_{JJ}))$. The fit was done for jets 
with $p^{min}_T=10\,\mathrm{GeV}$ and $p^{min}_T=25\,\mathrm{GeV}$, separately. 
\par The overall normalization factor cannot be studied since the CDF
collaboration 
did not determine the luminosity for the measurement. On the other hand, the relative normalization between 
the inclusive and exclusive production is a useful information. 
The relative normalization allows to make predictions for higher $p_T$
jets or for LHC energies for instance.  For this sake, the relative normalizations 
$r^{\mathrm{EXC}/\mathrm{INC}}$ should not vary much 
between the two $p^{min}_T$ measurements. 
Results are summarized in Table~\ref{TabRelNorm}. We give the inclusive and 
$\sigma^{\mathrm{INC}}$ and the exclusive cross sections 
$\sigma^{\mathrm{EXC}}$, obtained directly from
the models, and the relative scale factor needed to describe the CDF data 
to be applied to the exclusive contribution only. 
Whereas the relative normalization changes as a function $p_{T}^{min}$ by
an order of magnitude for the exclusive BL model, it tends to be rather stable for the 
KMR model (the uncertainty on the factor 2.5 might be relatively large
since we do not have a full
simulation interface and the simulation effects tend to be higher at low
jet transverse momentum). Finally, in Fig. \ref{FigAll10} and \ref{FigAll25}, 
the fitted distributions are depicted for $p_{T}^{min}=10,25\,
\mathrm{GeV}$ jets, respectively. 

\begin{table}[h]
\begin{tabular}{|ccc|c|c|c||c|c|c|}
\hline
 \multicolumn{3}{|c|}{contributions}     & $r^{\mathrm{EXC/INC}}(10)$     &  $\sigma^{\mathrm{INC}}(10)[\mathrm{pb}]$  &  $\sigma^{\mathrm{EXC}}(10)[\mathrm{pb}]$ & $r\mathrm{^{EXC/INC}}(25)$  &  $\sigma^{\mathrm{INC}}(25)[\mathrm{pb}]$  &  $\sigma^{\mathrm{EXC}}(25)[\mathrm{pb}]$  \\
\hline
FM &+& KMR        & 2.50            &   1249                        & 238        & 1.0      &   7.39   &  3.95             \\
FM &+& BL exc     & 0.35           &   1249                        & 1950          & 0.038    &   7.39   &  108            \\
BL inc &+& BL exc & 0.46           &   2000                        & 1950         & 0.017    &   40.6   &  108              \\
\hline
\end{tabular}
\caption{ Cross sections for inclusive diffractive production $\sigma^{\mathrm{INC}}$, exclusive
 cross section $\sigma^{\mathrm{EXC}}$ to be rescaled with a relative additional normalization between inclusive and 
exclusive events $r^{\mathrm{EXC/INC}}$
for $p_T>10\,\mathrm{GeV}$ and $p_T>25\,\mathrm{GeV}$ jets and for different models
(see text). Note that the fit to the data is parametriezed as $N(\sigma^{\mathrm{INC}}(R_{JJ}) + r^{\mathrm{EXC/INC}}\sigma^{\mathrm{EXC}}(R_{JJ}))$.} 
\label{TabRelNorm}
\end{table}

\par
\begin{figure}
\includegraphics[width=\picwidth]{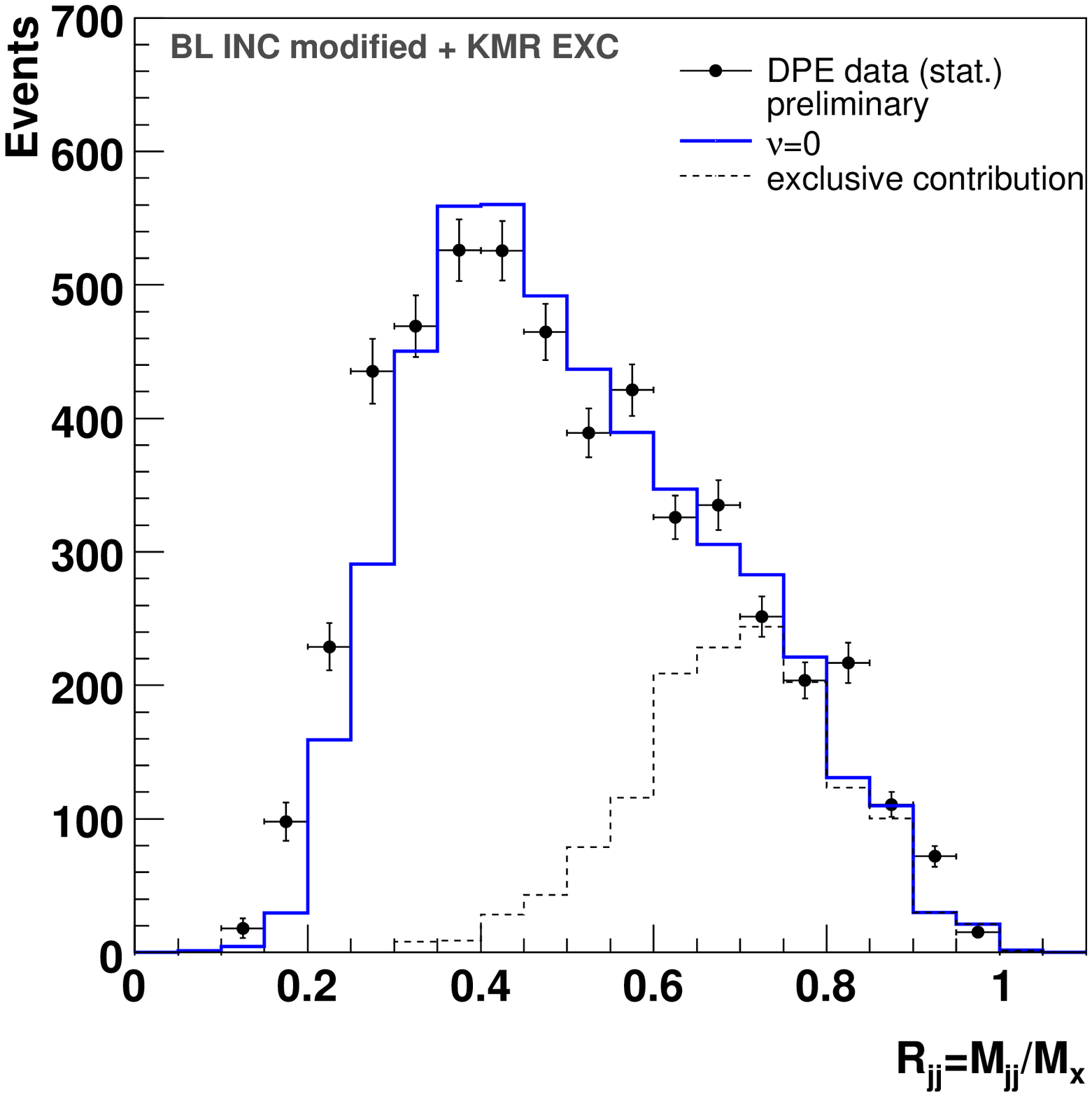}
\includegraphics[width=\picwidth]{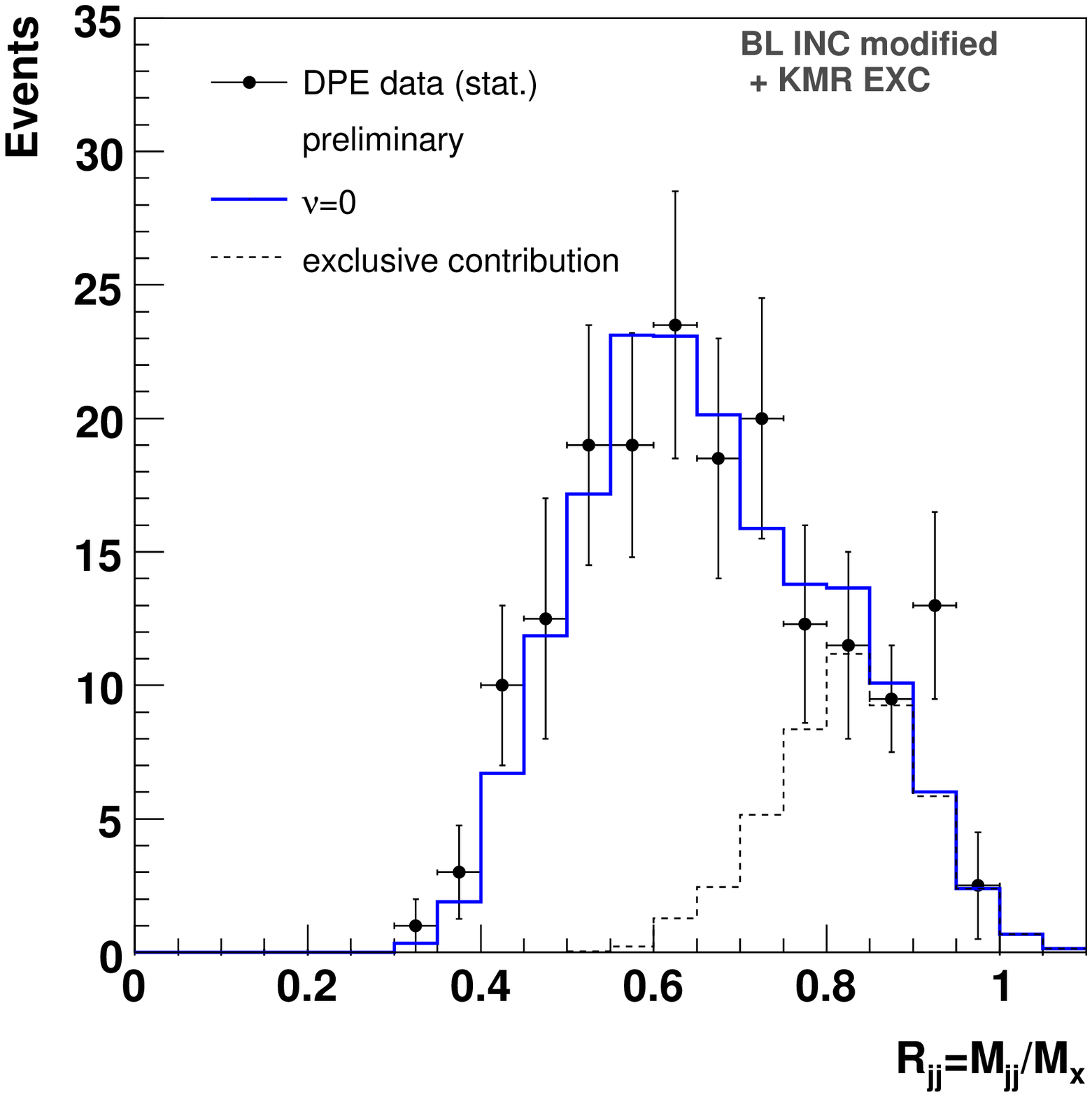}
\caption{Dijet mass distribution at the Tevatron calculated with the "modified" parton densities (see text) for 10\,GeV (left) and 25\, GeV (right) jets, KMR exclusive model included. }
\label{FigBLnb}
\end{figure}

 The Tevatron data are well described by the combination of FM and KMR model. We attribute
the departure from the smooth distribution of the data to the imperfection of our fast  simulation interface. On the 
contrary, the BL inclusive model is disfavoured because it fails to describe the low $R_{JJ}$ region.
It is due to the $\beta_i$ factor in the parton density $f_{i/\mathbb{P}}(\beta_i)$ used by the BL inclusive
model (see footnote 1 where the variables are defined) that the $R_{JJ}$ 
distribution is shifted towards higher values. This factor was
introduced to maintain the correspondence between the inclusive and exclusive 
model in the limit
$f_{i/\mathbb{P}}(x_i)\rightarrow \delta(x_i)$. On the contrary, this assumption
leads to properties in contradiction with CDF data.
Using the BL inclusive model without this additional normalization factor 
leads to a DMF which is in fair
agreement with data. Indeed, we show in Fig. \ref{FigBLnb} the predictions 
of  the ``modified" model (i.e. defined as 
$f_{i/\mathbb{P}}(\beta_i)\equiv G_{i/\mathbb{P}}(\beta_i)$) for $p_T>10$\,GeV and $p_T>25$\,GeV jets. We see that the
low $R_{JJ}$ region is described well and that fitting 
the prediction of the exclusive KMR model with the BL inclusive model 
yields roughly the same amount of exclusive events 
as using the factorable models. The
BL inclusive model will be revised to take these effects into account. We will
not mention further this "modified" version of the BL inclusive model since
it gives similar results as the factorable models.

\par The exclusive BL model leads to a quite reasonable description of the DMF shape for both $p^{min}_T$ cuts in
combination  with FM, however, it fails to grasp the shape of the exclusive cross section measured as a 
function of the jet minimal 
transverse momentum $p_T^{min}$. To illustrate this, we present the CDF data for exclusive cross section corrected for 
detector effects compared with the predictions of both exclusive models 
after applying the same cuts as in the CDF measurement,
namely: $p^{jet1,2}_T>p_T^{min}$, $|\eta^{jet1,2}|<2.5$,
$3.6<\eta_{gap}<5.9$, $0.03 < \xi_{\bar{p}}<0.08$. The BL exclusive model 
shows a much weaker $p_T$ dependence than the KMR
model and is in disagreement with data. \footnote{Let us note that 
the cross section of exclusive events measured by the CDF collaboration 
is an indirect measurement since it was obtained by subtracting the inclusive
contribution using an older version of the gluon density in the pomeron measured
at HERA. In that sense, the contribution of exclusive events using the newest
gluon density from HERA might change those results. However, as we noticed,
modifying the gluon density even greatly at high $\beta$ by multiplying
the gluon distribution by $(1-\beta)^{\nu}$ does not change the amount of
exclusive events by a large factor, and thus does not modify the indirect
measurement performed by the CDF collaboration much.} 

To finish the discussion about the pomeron like models, it is worth mentioning
that these results assume that the survival probability has no strong dependence
on $\beta$ and $\xi$. If this is not the case, we cannot assume that the shape
of the gluon distribution as measured at HERA could be used to make predictions
at the Tevatron. However, this is a reasonable assumption since the survival
probability is related to soft phenomena occuring during hadronisation effects
which occur at a much longer time scale than the hard interaction. In other
words, it is natural to suppose that the soft phenomenon will not be influenced
by the hard interaction.

\begin{figure}
\includegraphics[width=\picwidth]{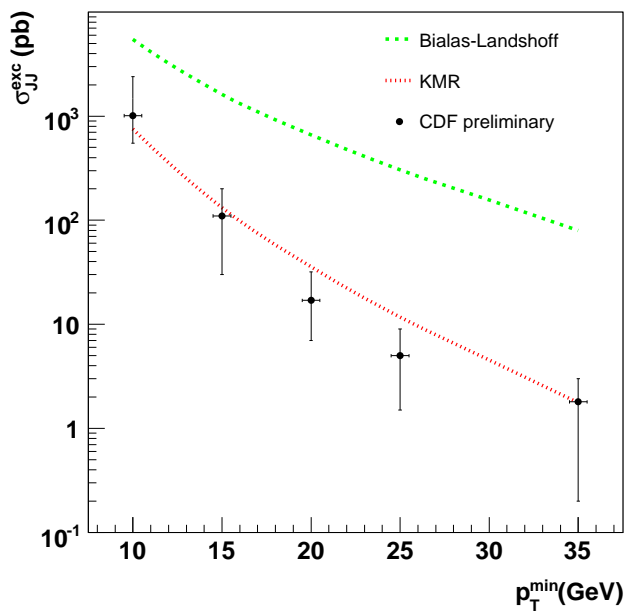}
\caption{Exclusive cross section as a function of the minimal transverse jet momentum $p^{min}_T$ measured by the 
CDF collaboration and compared to the prediction of the KMR and BL exclusive 
models. We note that the BL model overshoots the CDF measurement while the KMR
model is in good agreement.}
\label{FigSigmaEXC}
\end{figure}

\begin{figure}[h]
\includegraphics[width=\picwidth]{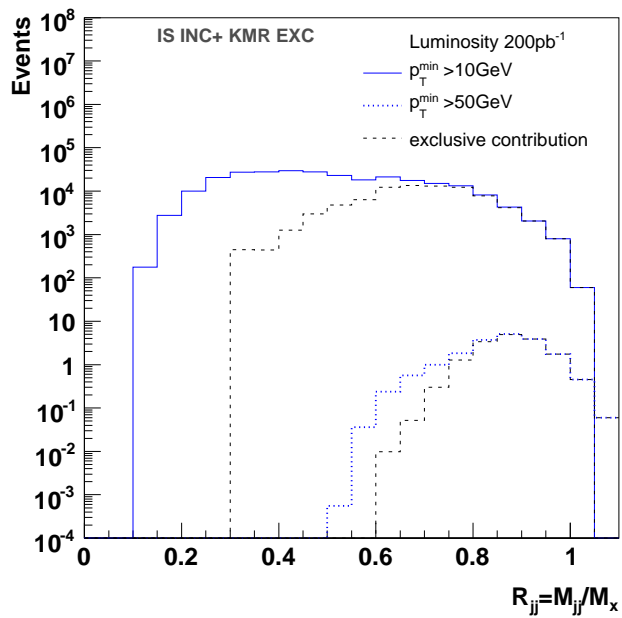}
\caption{Dijet mass fraction for two values of minimal transverse jet momentum
$p^{min}_T$. We note that the relative exclusive contribution is higher
at high $p^{min}_T$.}
\label{FigDMFpt}
\end{figure}

\begin{figure}
\includegraphics[width=\picwidth]{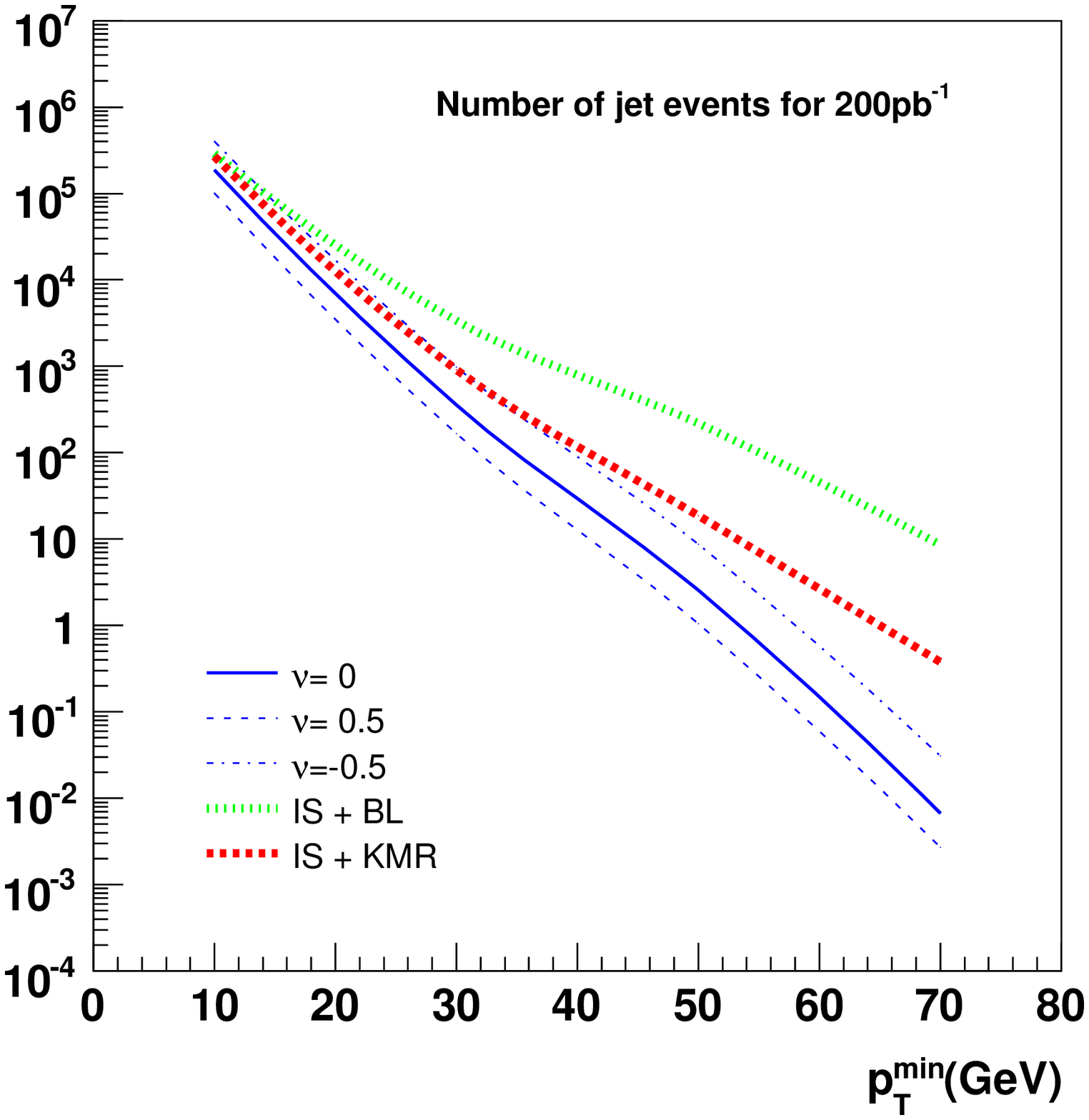}
\includegraphics[width=\picwidth]{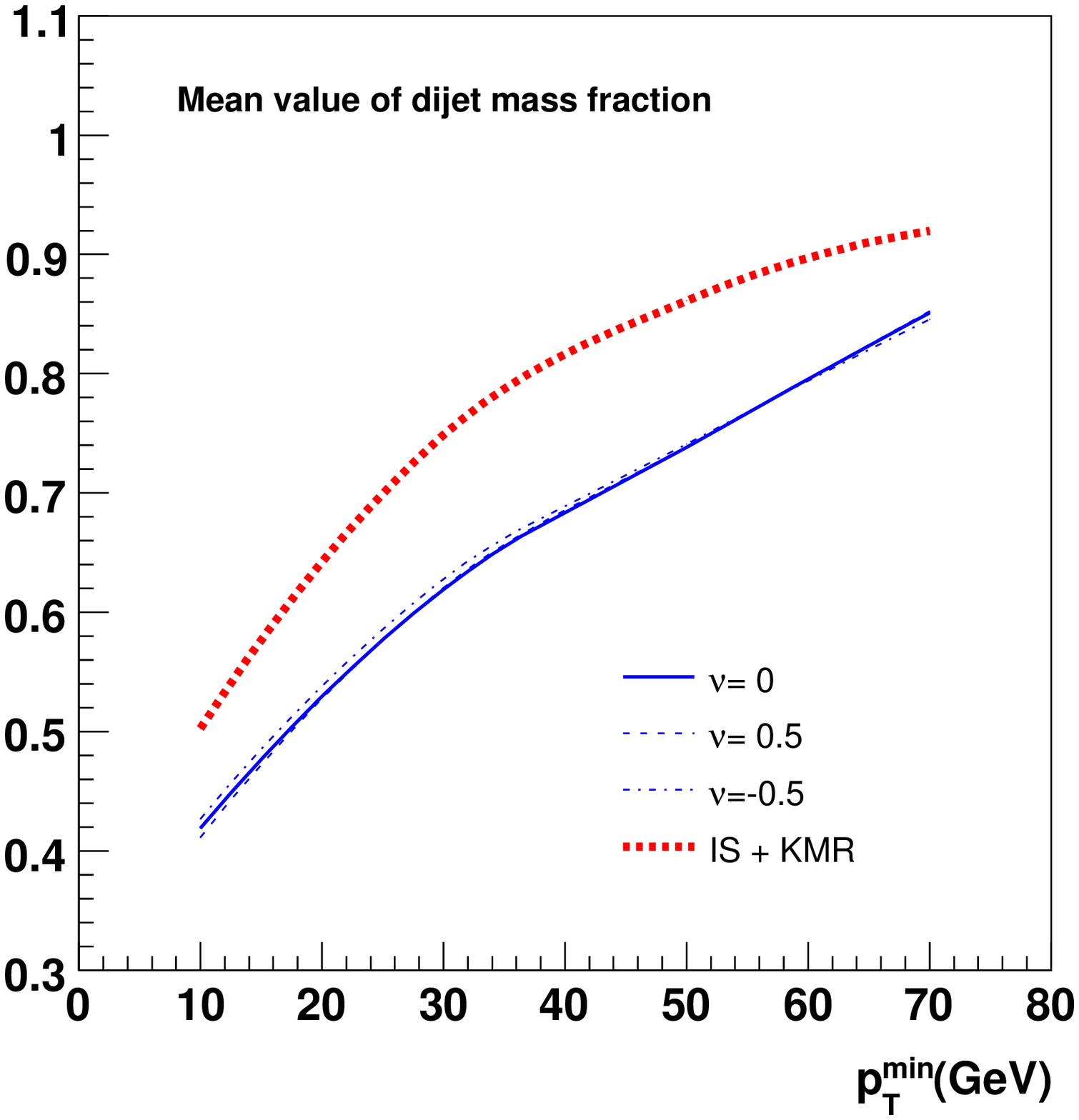}
\caption{Number of jet events and mean of the  dijet mass fraction as a 
function of the minimal jet $p^{min}_T$. We note that the ideal value of 
$p^{min}_T$ to enhance the exclusive contribution is of the order of 30-40 GeV
which leads to a high enough production cross section as well as a large effect of the
exclusive contribution on the dijet mass fraction.}
\label{FigNDPmean}
\end{figure}
\subsection{Prospects of future measurements at the Tevatron}

In this section, we list some examples of observables which could be used 
to identify better the exclusive contribution in DMF measurements at the Tevatron. We present the 
prediction as a function of the minimal transverse momentum of the two leading jets $p_T^{min}$. Since
the BL inclusive model does not describe the DMF at low $R_{JJ}$, we choose to
show only the FM prediction in combination with both, KMR and BL exclusive models.

\par The same roman pot acceptance and restriction cuts as in the CDF 
measurement were used, specifically, $0.01<\xi_{\bar{p}}<0.12$, $p_T^{jet1,2}> p_T^{min}$, $|\eta^{jet1,2}|<2.5$, $3.6<|\eta_{gap}|<5.9$. Moreover, we adopted a normalization between
inclusive and exclusive events as obtained for the $p_T>25\,\mathrm{GeV}$ analysis in the previous section because we
are less sensitive to the imperfections of the fast simulation interface for higher $p_T$ jets. Fig. \ref{FigDMFpt}
illustrates the appearance of DMF for two separate values of minimum jet $p_T^{min}$. The character of the distribution is clearly 
governed by exclusive events at high $p^{min}_T$. 
\par Fig. \ref{FigNDPmean} shows the rate of DPE events. 
In addition to the curves denoting inclusive contribution with the varied gluon density
for $\nu=-0.5,0, 0.5$, the full contribution for both exclusive models is shown. 
For the FM model which is in better consistency with accessible data, the measurement of the
DPE rate does not provide an evident separation of exclusive contribution from 
the effects due to the pomeron uncertainty since the noticable difference appears when the cross
sections are too low to be observable.
It is possible, however, to examine the mean of the DMF distribution. As seen in Fig. \ref{FigNDPmean},
this observable disentangles well the exclusive production with the highest effect between 30 and $40\,\mathrm{GeV}.$

\par It needs to be stressed that even though  we obtain a hint in understanding the exclusive
production phenomena at the Tevatron, the final picture cannot be drawn before precisely measuring the 
structure of the pomeron. For this purpose the DMF or the DPE rate are not suitable at the Tevatron. In the former, 
there is no sensitivity to the high $\beta$ gluon variation, whereas 
in the latter, the gluon variation and the exclusive contribution cannot be 
easily separated. The way out is to perform QCD fits of the pomeron structure
in gluon and quark for data at low
$R_{JJ}$
where the exclusive contribution is negligible. Another possibility is to 
perform silmutaneously the global fits of pomeron structure functions using 
DGLAP evolution and of the exclusive production. 
\par
A final important remark is that this study was assuming pomeron like models for
inclusive diffraction. It is worth studying other models like Soft color
interaction processes and find out if they also lead to the same conclusion
concerning the existence of exclusive events.

\subsection{Soft color interaction model}

\begin{figure}
\parbox{\picwidth}{
\includegraphics[width=\picwidth]{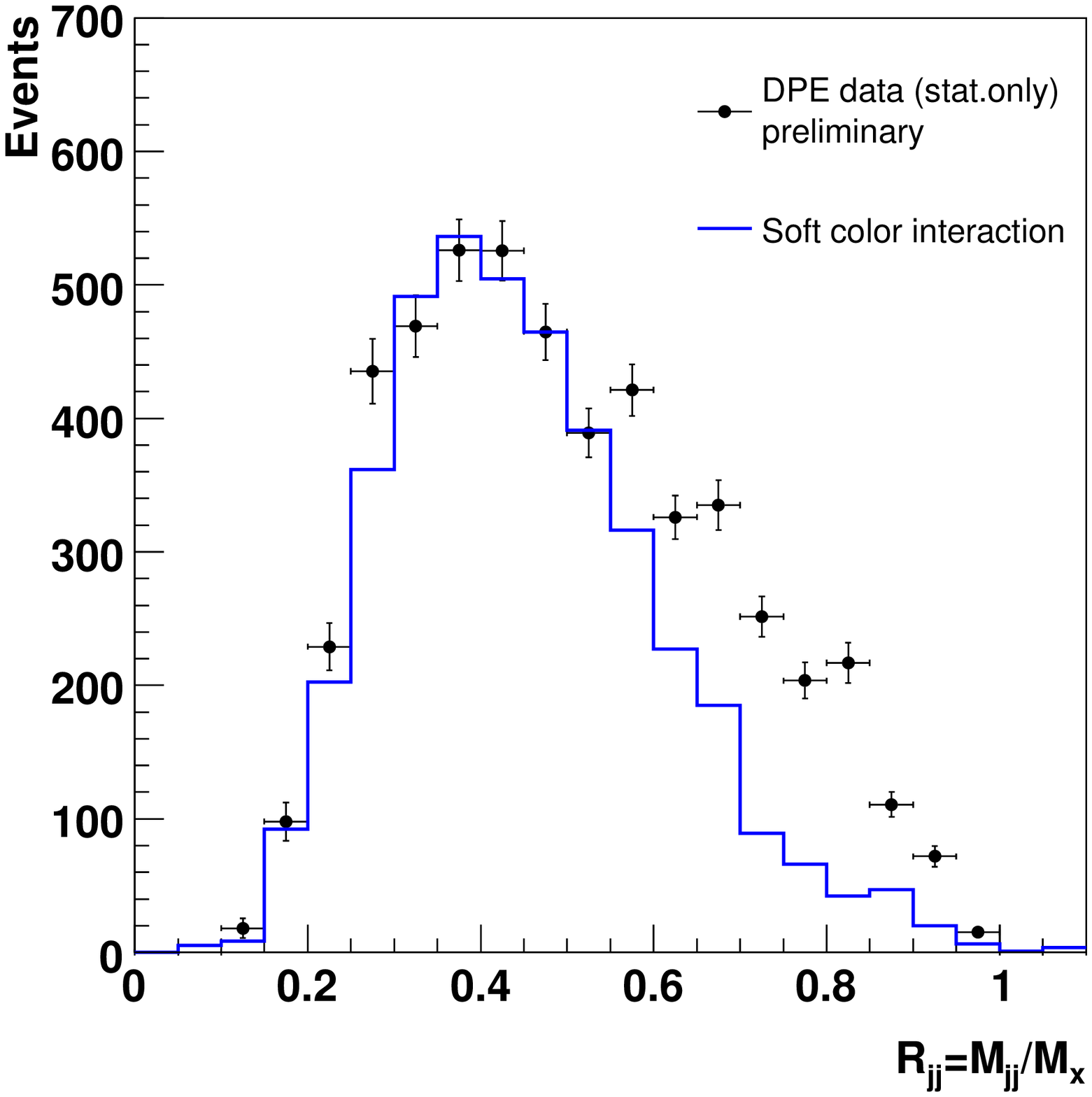}
}
\parbox{\picwidth}{\includegraphics[width=\picwidth]{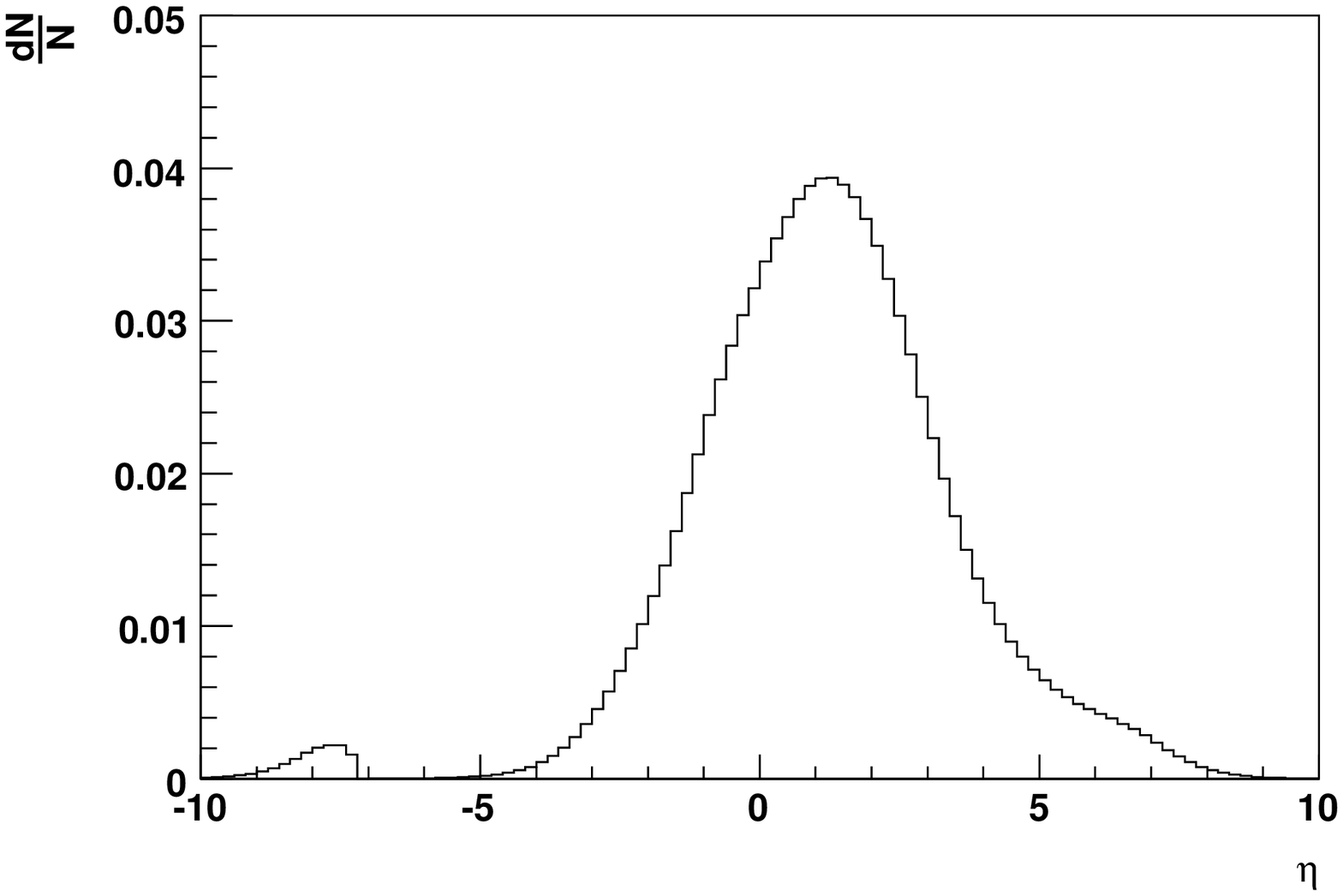}
}
\caption{Dijet mass fraction at the Tevatron for jets $p_T>10\,\mathrm{GeV}$ 
(left) and the $\eta$ distribution of produced particles (right) for the
Soft color interaction model.}
\label{Figsciflow}
\end{figure}

\begin{figure}
\includegraphics[width=\picwidth]{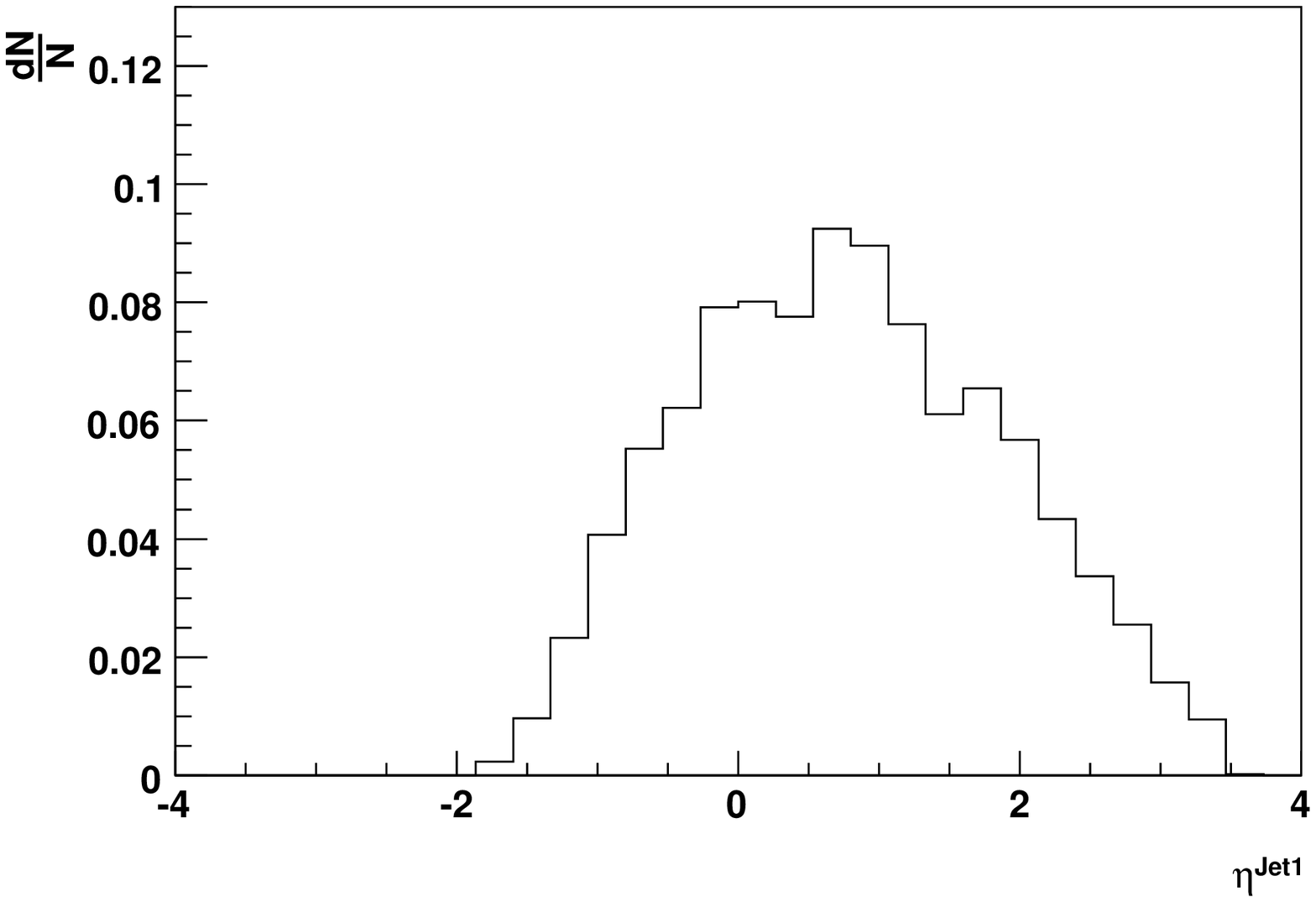}
\includegraphics[width=\picwidth]{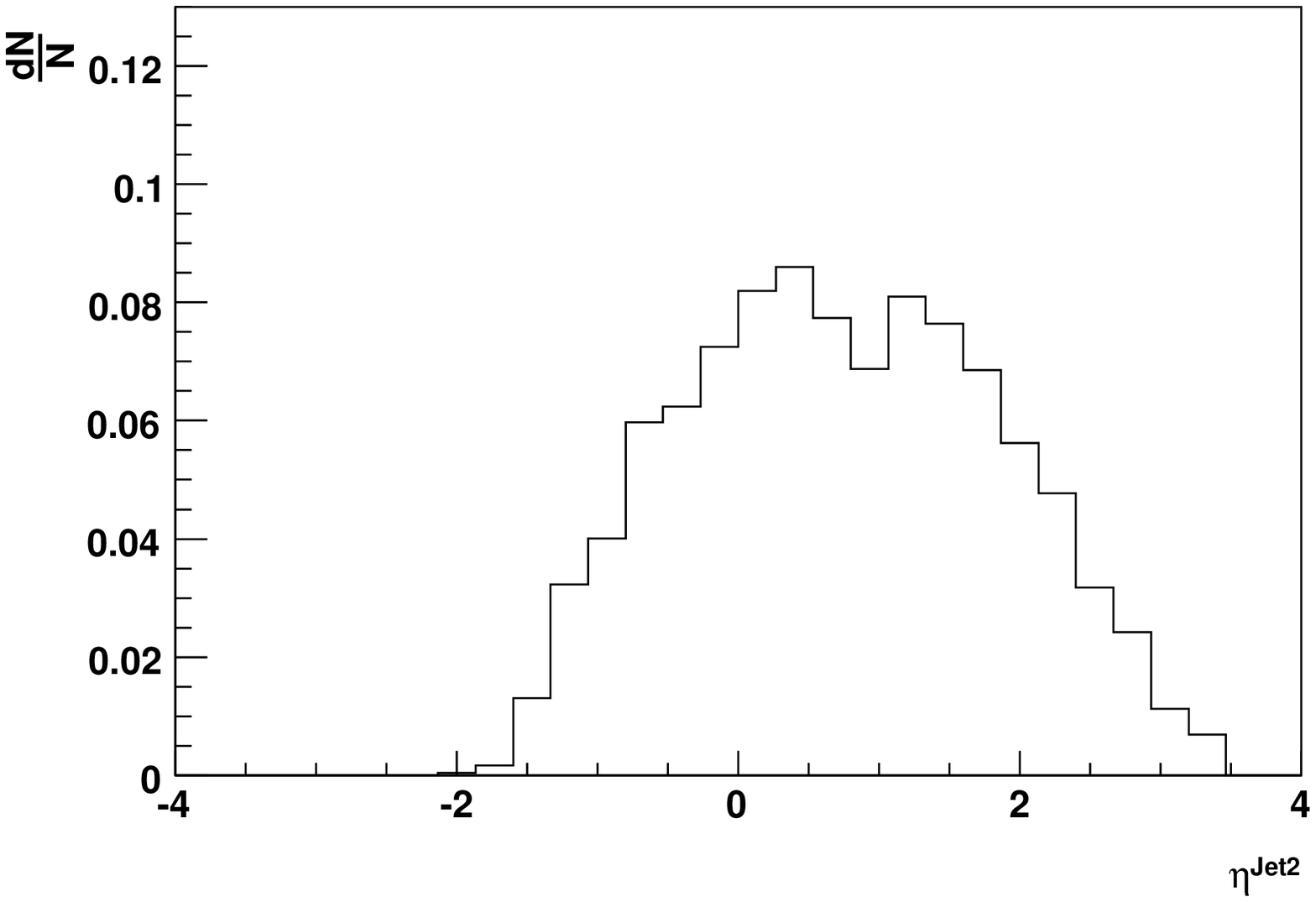}
\caption{Rapidity distribution of a leading jet (left) and a second leading
jet (right) in the SCI model when calculating dijet mass fraction.}
\label{Figscijets}
\end{figure}

\begin{figure}
\includegraphics[width=\picwidth]{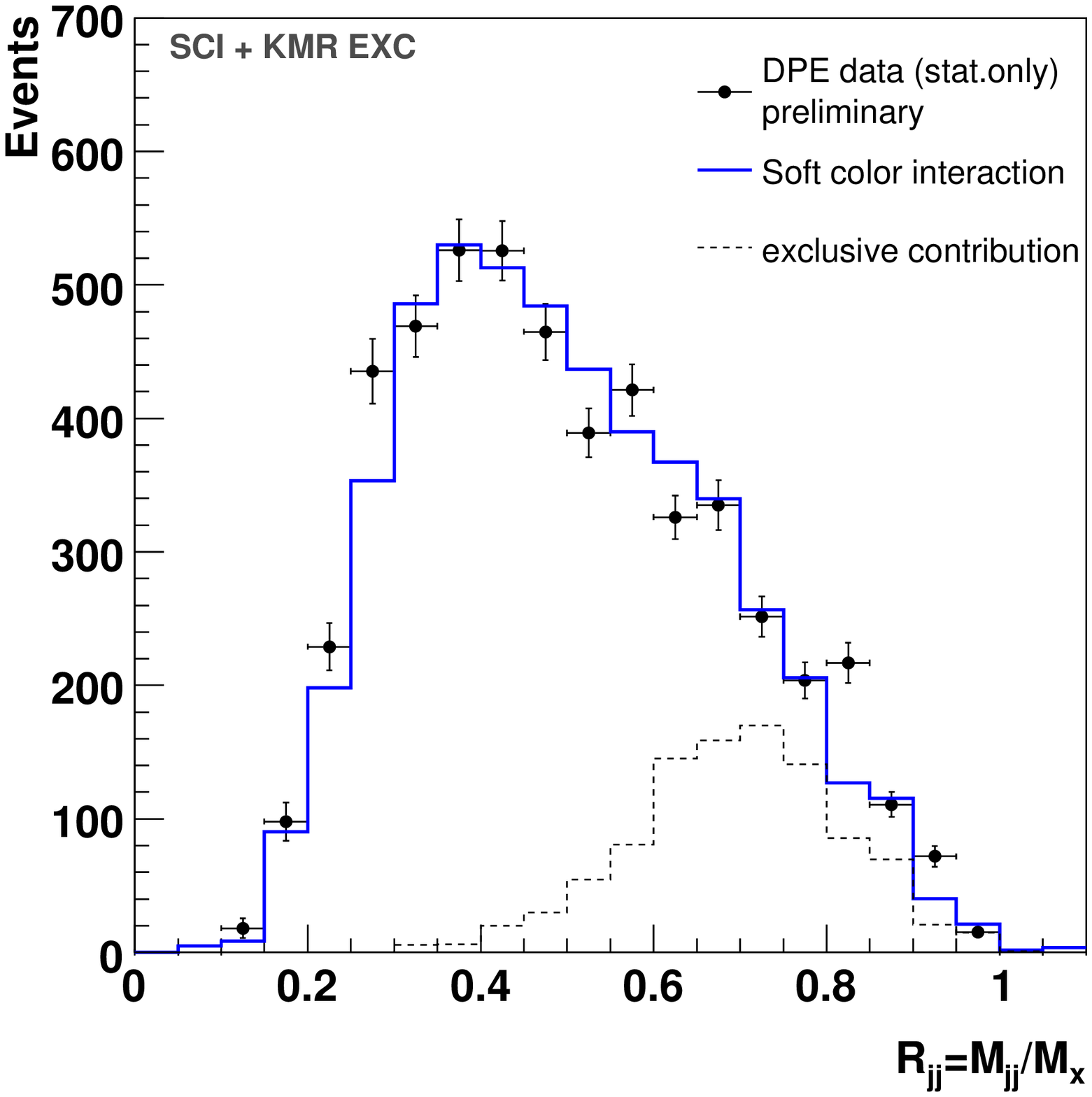}
\includegraphics[width=\picwidth]{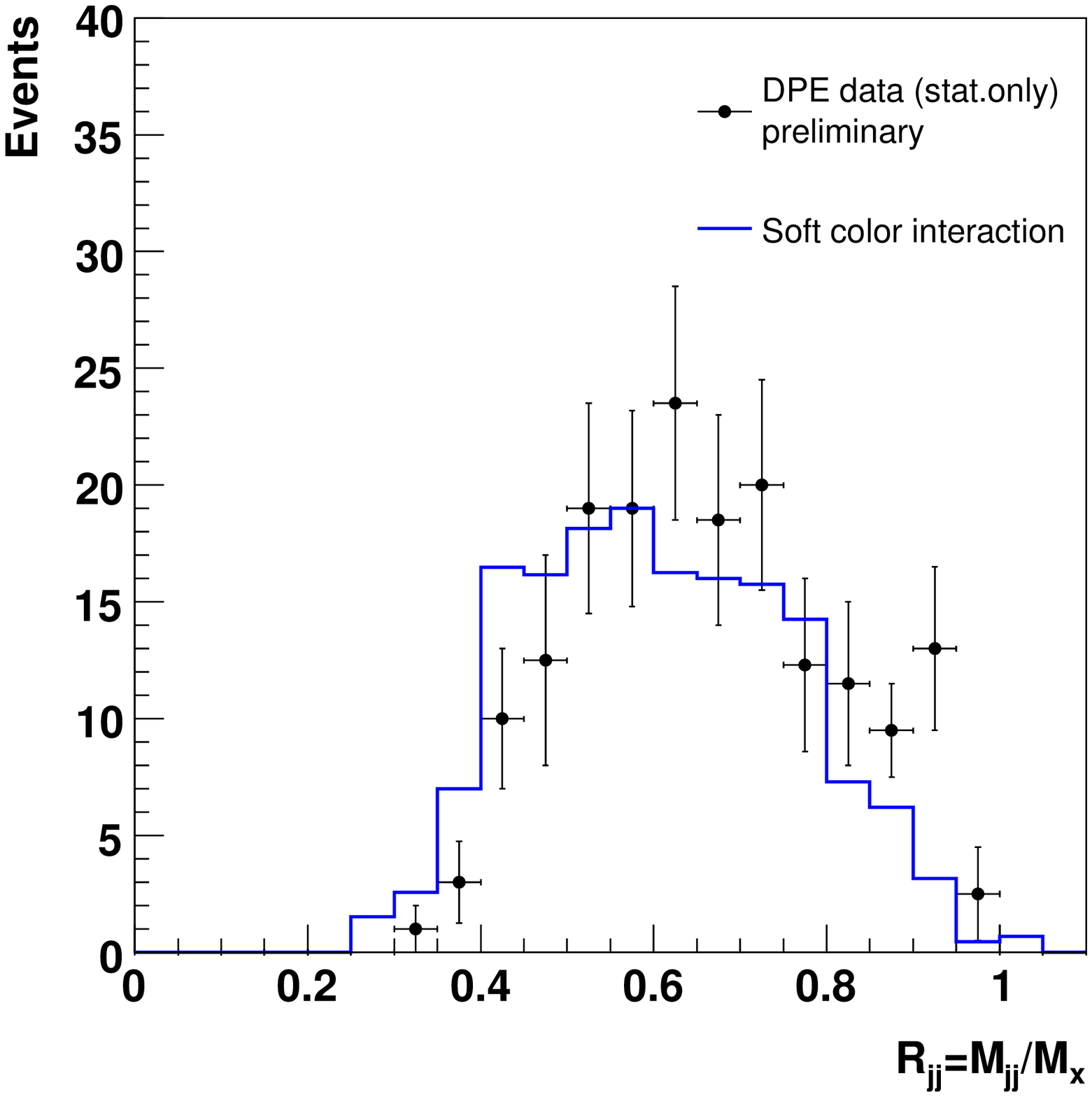}
\caption{Dijet mass fraction at the Tevatron for jets $p_T>10\,\mathrm{GeV}$ 
for the SCI model and KMR exclusive
model (left), and for jets $p_T>25\,\mathrm{GeV}$  for the  SCI model only (right).}
\label{dmfsci25}
\end{figure}

The Soft color interaction model uses different approach to explain diffractive
events. In this model, diffraction is due to special color rearrangement
in the final state as we mentionned earlier. It is worth noticing that in this
model, the CDF data are dominated by events with tagged antiproton on the $\bar{p}$ ($\eta_{\bar{p}}<0$) side
and a rapidity gap on the $p$ side. In other words, in most of the events,
there is only one single antiproton in the final state accompanied by a bunch of 
particles (mainly pions) flowing into the beam pipe. This is illustrated in Fig. \ref{Figsciflow}
right which shows the rapidity distribution of produced particles and we
notice the tail of the distribution at high rapidity. 
We should not omit to mention that on the other hand, the probability to get two protons intact (which is important for the double tagged events) is
in SCI model extremly small.
\par
After applying all CDF cuts mentioned above, the comparison between SCI and 
CDF data on $R_{JJ}$ is shown in Figs. \ref{Figsciflow} (left) and \ref{dmfsci25}. 
Whereas it is not possible to
describe the full dijet mass fraction for a jet with $p_T>10\,$GeV, it is noticeable
that the exclusive contribution is found to be lower than in the case of the
pomeron inspired models. Indeed, performing the same independent fit of SCI 
and KMR exlusive contribution one finds that only 70\,\% of 
the exclusive contribution needed
in case of pomeron inspired models is necessary to describe the data. 
For jets with $p_T>25\,$GeV, no additional exclusive contribution is needed 
to describe the 
measurement which can be seen in Fig.~\ref{dmfsci25}.
Since most events are asymmetric in the sense that only the antiproton is strictly intact 
and on the other side, there is a flow of particles in the beam pipe, it is
worth studying the rapidity distribution of jets for this model. The results are
shown in Fig. \ref{Figscijets}. We note that the rapidity distribution is
boosted towards high values of rapidity and not centered around zero like for
pomeron inspired models and CDF data. Moreover, the cross section  for
$p_T>10\,$GeV jets is in the SCI model $\sigma^{\mathrm{SCI}}=167\,$pb,
about only 
13\% of the cross section predicted by the pomeron inspired models 
which however give a correct prediction of a large range of observables including 
DPE cross sections. Therefore, such properties disfavour the SCI model.
However, it would be worth studying and modifying the SCI model 
since the probability to observe two protons in the final state (and/or two gaps) 
should be higher than the square probability of observing one proton (and/or one gap) only
(single diffraction) as it was seen by the CDF collaboration \cite{cdfprob}. The model
needs to be adjusted to take this into account and than it would be interesting
to see the
impact on the dijet mass fraction and the existence of exclusive events.

\section{Dijet mass fraction at the LHC}
It was suggested that exclusive production at the LHC could be used to study the
properties of a specific
class of centrally produced objects like Higgs bosons. However, it relies on 
many subtleties such as a good 
understanding of the inclusive production. The perturbative nature of the 
diffractive processes results
in the factorization of the cross section to a regge flux and pomeron structure
functions, while factorization breaking appears via the survival probability only.
The gluon density in the pomeron is of most important matter, 
since its value at high momentum  fraction will control the background to 
exclusive DPE, and the pomeron flux and the survival probability factor will have to be measured at the LHC to make reliable predictions. 

\begin{figure}[h]
\begin{center}
\includegraphics[width=\picwidth]{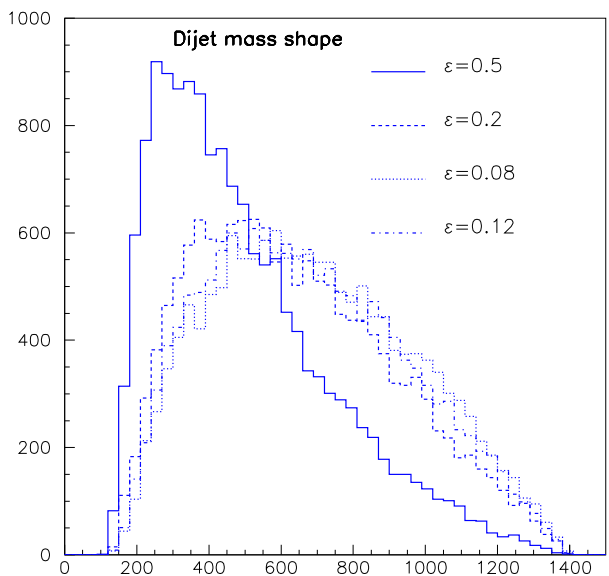}
\caption{Sensitivity of the dijet mass fraction to different values of the
pomeron intercept $\alpha_\mathbb{P} = 1 + \epsilon$. }
\label{FigInterceptLHC}
\end{center}
\end{figure}

The flux depends on the pomeron intercept $\alpha_\mathbb{P}$ whose impact on 
the DMF distribution for LHC energies
is shown in Fig.~\ref{FigInterceptLHC}. The pomeron intercept is parametrized 
as $\alpha_\mathbb{P}=1+\epsilon$ and the prediction is made for four values 
of $\epsilon=0.5, 0.2, 0.12, 0.08$. The updated HERA pomeron structure function analysis \cite{pdfs} suggests
that the ``hard pomeron" intercept value is close to $\alpha_\mathbb{P}=1.12$. Nevertheless, new QCD fits using
single diffractive or double pomeron exchange data will have to be performed to fully constrain the parton
densities and the pomeron flux at the LHC.

\par We also give the dependence of the DMF on jet $p_T$ at the LHC. 
DPE events in this analysis were selected applying the roman pot acceptance on 
both 
sides from the interaction point, and using a fast simulation of the CMS detector 
\cite{cmssim} (the results would be similar using the ATLAS simulation)
and asking two 
leading jets with $p_T>=100,200,300,400\,\mathrm{GeV}$.
We have disfavored the predictions of the BL exclusive model at the Tevatron.  The BL exclusive shows
weak $p_T$ dependence which makes the model unphysical for LHC energies since it predicts 
cross sections even higher than the inclusive ones. We therefore focus on the predictions of FM and 
KMR models, only. As in the previous sections, we also include a study of the uncertainty on the gluon 
density enhancing the high $\beta$ gluon with a factor $(1-\beta)^{\nu}$.

\begin{figure}[h]
\begin{center}
\epsfig{file=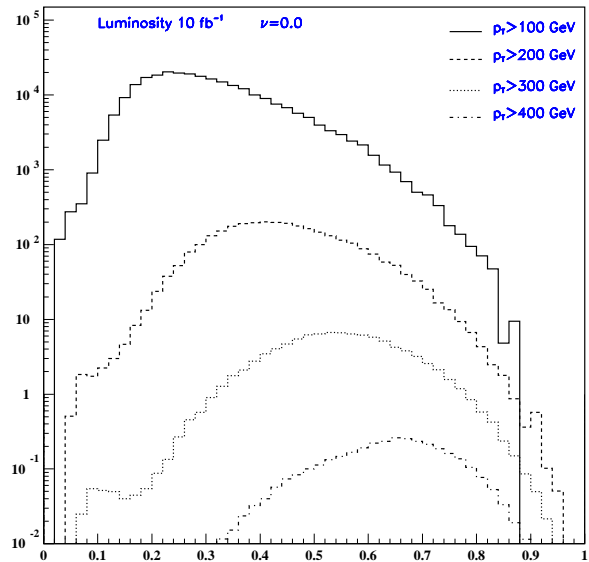,width=\picwidth}
\caption{Dijet mass fraction at the LHC as a function of jet minimal transverse momentum $p^{min}_T$, FM inclusive model.}
\label{FigDMFLHC}
\end{center}
\end{figure} 

\begin{figure}[h]
\begin{center}
\epsfig{file=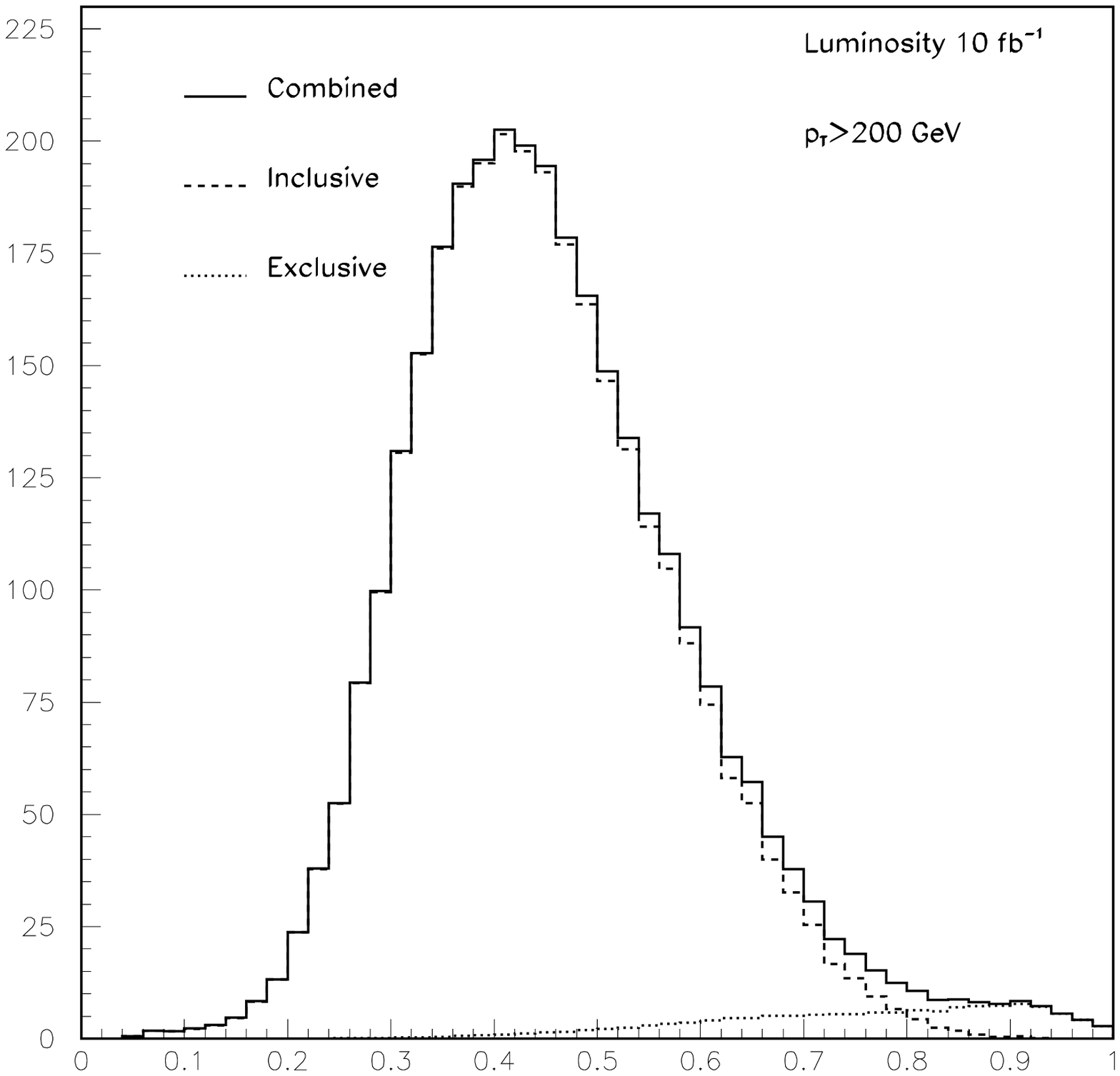,width=\picwidth} 
\epsfig{file=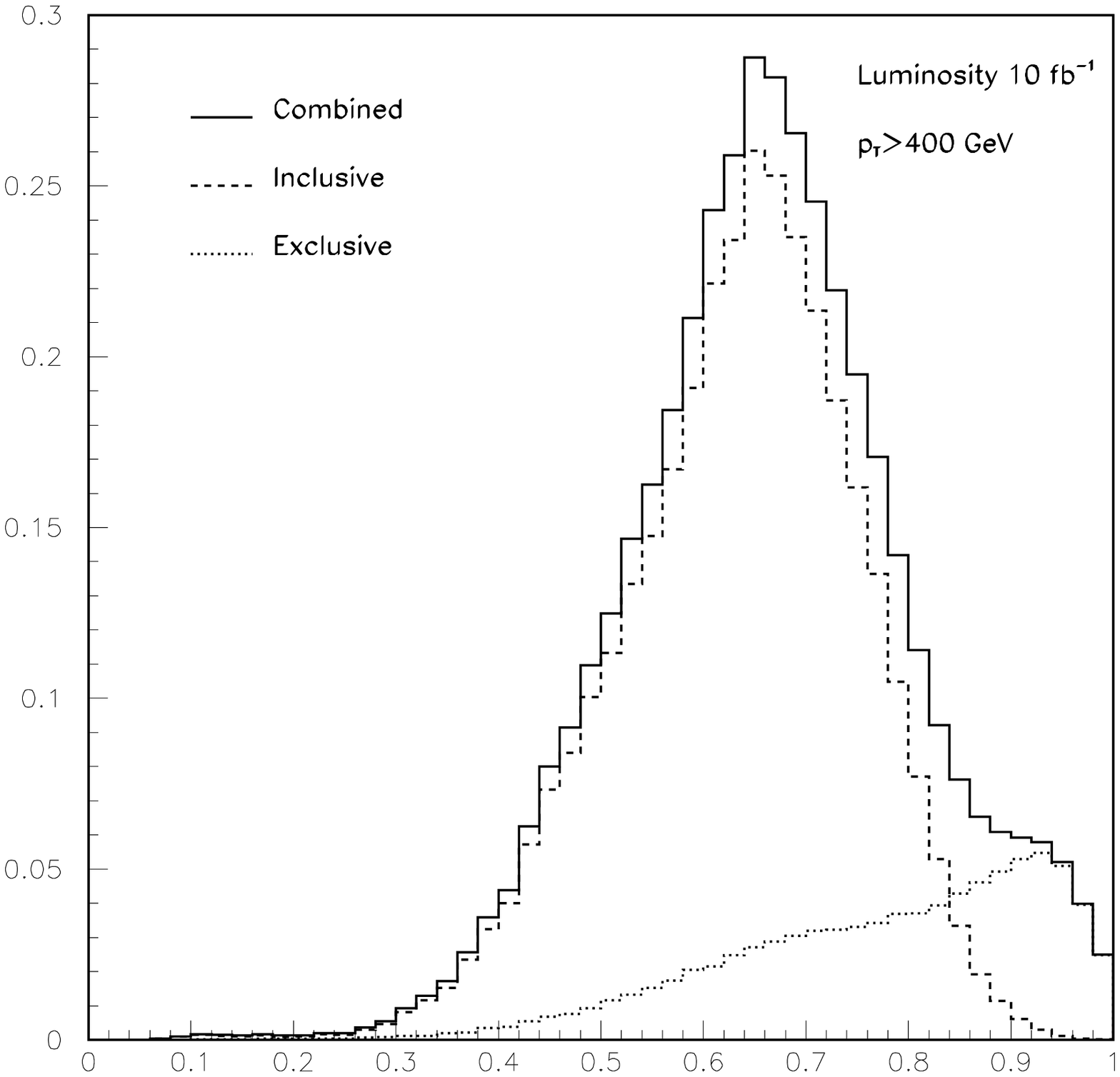,width=\picwidth} 
\caption{Dijet mass fraction at the LHC for jets $p_T>200\,\mathrm{GeV}$ and $p_T>400\,\mathrm{GeV}$, respectively, FM inclusive + KMR exclusive models.}
\label{FigDMFexcLHC}
\end{center}
\end{figure} 

\begin{figure}[h]
\begin{center}
\epsfig{file=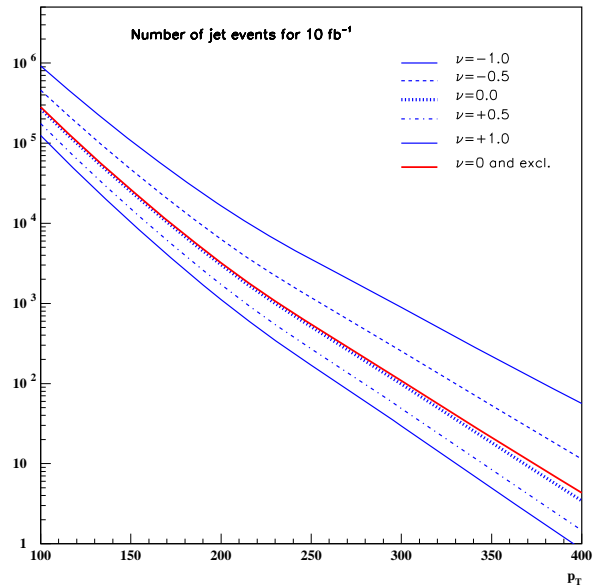,width=\picwidth}
\caption{Number of DPE events at the LHC
as a function of minimal transverse momentum $p^{min}_T$ of two leading jets. FM inclusive + KMR exclusive models. The
gluon variation is displayed for different $\nu$ values.}
\label{FigNDPEptLHC}
\end{center}
\end{figure}

\begin{figure}[h]
\begin{center}
\epsfig{file=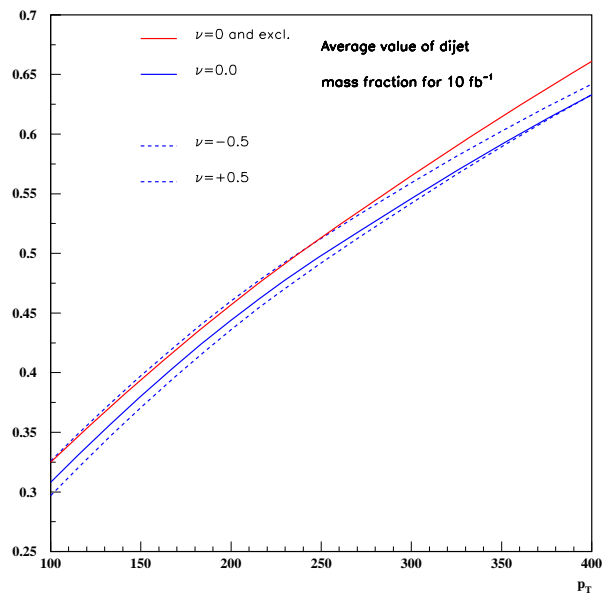,width=\picwidth}
\caption{Average value of the dijet mass fraction
as a function of minimal transverse momentum $p^{min}_T$ of the leading jets. Exclusive contribution and different values of $\nu$ 
are shown. FM + KMR models.}
\label{FigMeanLHC}
\end{center}
\end{figure} 

\clearpage
\par The dijet mass fraction as a function of different $p_T$ is visible in Fig.~\ref{FigDMFLHC}. The
exclusive contribution manifests itself as an increase in the tail of the distribution which can be seen for $200\,\mathrm{GeV}$ 
jets (left) and $400\,\mathrm{GeV}$ jets (right), respectively in
Fig. \ref{FigDMFexcLHC}. Exclusive production slowly 
turns on with the increase of the jet $p_T$ which is demonstrated in Fig. 
\ref{FigNDPEptLHC} where the number of expected DPE events is shown. However, 
with respect to the uncertainty on the gluon density this appearance is 
almost negligiable. One can use the average position
of the DMF as a function of the minimal jet transverse momentum $p_T^{min}$
to study the presence of the exclusive contribution,
see Fig. \ref{FigMeanLHC}.
This is true especially for high $p_T$ jets.

\par The exclusive production at the LHC plays a minor role for low $p_T$ jets. Therefore, measurements e.g for $p_T<200\,\mathrm{GeV}$ where the inclusive production is dominant could be used 
to constrain the gluon density in the pomeron. Afterwards, one can look in the high $p_T$ jet region to extract the exclusive 
contribution from the tail of the DMF.

\section{Conclusion}
The aim of this paper was to investigate whether we can 
explain the excess of events at the high dijet mass fraction measured 
at the Tevatron without the exclusive production. The result is actually 
two fold.
\par Concerning the pomeron induced models ("Factorized model" and Bialas-Landshoff 
inclusive models) we found that the uncertainty
on the high $\beta$ gluon density in the Pomeron has a 
small impact at high $R_{JJ}$. Therefore, an additional contribution 
is needed to describe the CDF data with these models. 
We examined the exclusive KMR model and Bialas-Landshoff exclusive model predictions
for the role of the additional contribution and found that the best descriprion 
of data is achieved by the combination of the Factorized inclusive model 
(or the modified inclusive Bialas-Landshoff one) and the KMR exclusive 
model. The exclusive contribution at the Tevatron can be magnified requesting higher 
$p_T$ jets  and studying specific observables like a mean of 
the dijet mass fraction, for example. Though, one of the limitations of 
using high $p_T$ jets is due to the rate of DPE events
which falls logarithmicaly allowing measurements for jets up to 
approximately $40\,\mathrm{GeV}$. The Bialas-Landshoff exclusive model seems 
to be disfavoured by  Tevatron
data since it shows a softer jet $p_T$ dependance and predicts unphysically 
large DPE rates at 
LHC energies. 
\par
In the case of the Soft color interaction model which is not based on pomeron exchanges,
the need to introduce an additional exclusive production is less obvious. For low $p_T$ jets
the amount of exclusive events to describe the data is smaller than in case of Factorized
model, but for high $p_T$ jets no additional contribution is necessary. This draws a new 
question: whether the double pomeron exchange events could be explained by 
a special rearrangement
of color only? The CDF data are in this model dominated by single
diffractive events. The probability of tagging two protons in the final state within this model
is very small, contradicting the CDF observation. So even though the SCI
model is not applicable for DPE events in the current state  it would be worth adjusting this model
to correctly predict the rate of double tagged events and study the model prediction of dijet mass fraction and other DPE induced processes. 
\par
Dijet mass fraction at the LHC could be used to select the exclusive events. 
Indeed, it is possible to study
jets with $p_T>200\,\mathrm{GeV}$ for instance, and to focus on events with DMF above 0.8 which
is dominated by exclusive production (see Fig.~\ref{FigDMFexcLHC}). However, 
as it was advocated earlier, a complete QCD analysis consisting of measuring 
the gluon density in the Pomeron (especially at high $\beta$) and study
the QCD evolution of exclusive events as a function of jet $p_T$ is needed
to fully understand the observables, and make predictions for
diffractive Higgs production and its background at the LHC as an example.

\section*{Acknowledgments}
The authors want to thank M. Boonekamp, R. Enberg, D. Goulianos, G. Ingelman, R. Pechanski 
and K. Tereashi for useful discussions and for
providing them the CDF data and roman pot acceptance.

\newpage

\section{Appendix}

Throughout the paper, we have purpously omitted a discussion of imperfections 
concerning the dijet mass fraction reconstruction within our framework, 
postponing it to this section. In this appendix, all calculations are done for jets with 
$p_T>10\,\mathrm{GeV}$.
\par 
\begin{itemize}
\item In our analysis, we define the dijet mass fraction as a ratio
of the two leading jet invariant mass $M_{JJ}$ to the central diffractive mass $M_X$.
The latter was determined using the momentum loss $\xi_{\bar{p}}$ measured in a roman pot 
on the antiproton side and the $\xi_p^{part}$ obtained from particles on the generator level,
such as $M_X=(s\xi_{\bar{p}}\xi^{part}_{p})^{1/2}$. In this case, we must ensure that all
of the produced diffractive energy $M_X$ is deposited into the central detector.
If this is not the case, our $M_X$ at generator level might be sensibly larger than
the one measured by the CDF collaboration. The energy 
flow of the particles on the generator level as a function of rapidity is 
shown in Fig. 
\ref{FigEnergy}, upper plot. The middle plot shows the energy flow weighted by the transverse momentum of the particle 
$E_T$. We see that most of the energy is deposited in the calorimeter region, i.e. for
$|\eta|<4$. In $\bar{p}$ tagged events, protons most frequently loose a smaller momentum 
fraction (roughly $\xi_p\sim0.025$) than the tagged antiproton for which the acceptance turns on
for $\xi_{\bar{p}}>0.035$. This can be seen from the $\xi_p$ population plot in 
the bottom of Fig. \ref{FigEnergy}. Thus,  a collision of more energetic pomeron from the 
antiproton side 
with a pomeron from the proton side is boosted 
towards the  $\bar{p}$ as it is seen on the energy flow distributions. 
\begin{figure}[h]
\includegraphics[width=\picwidth]{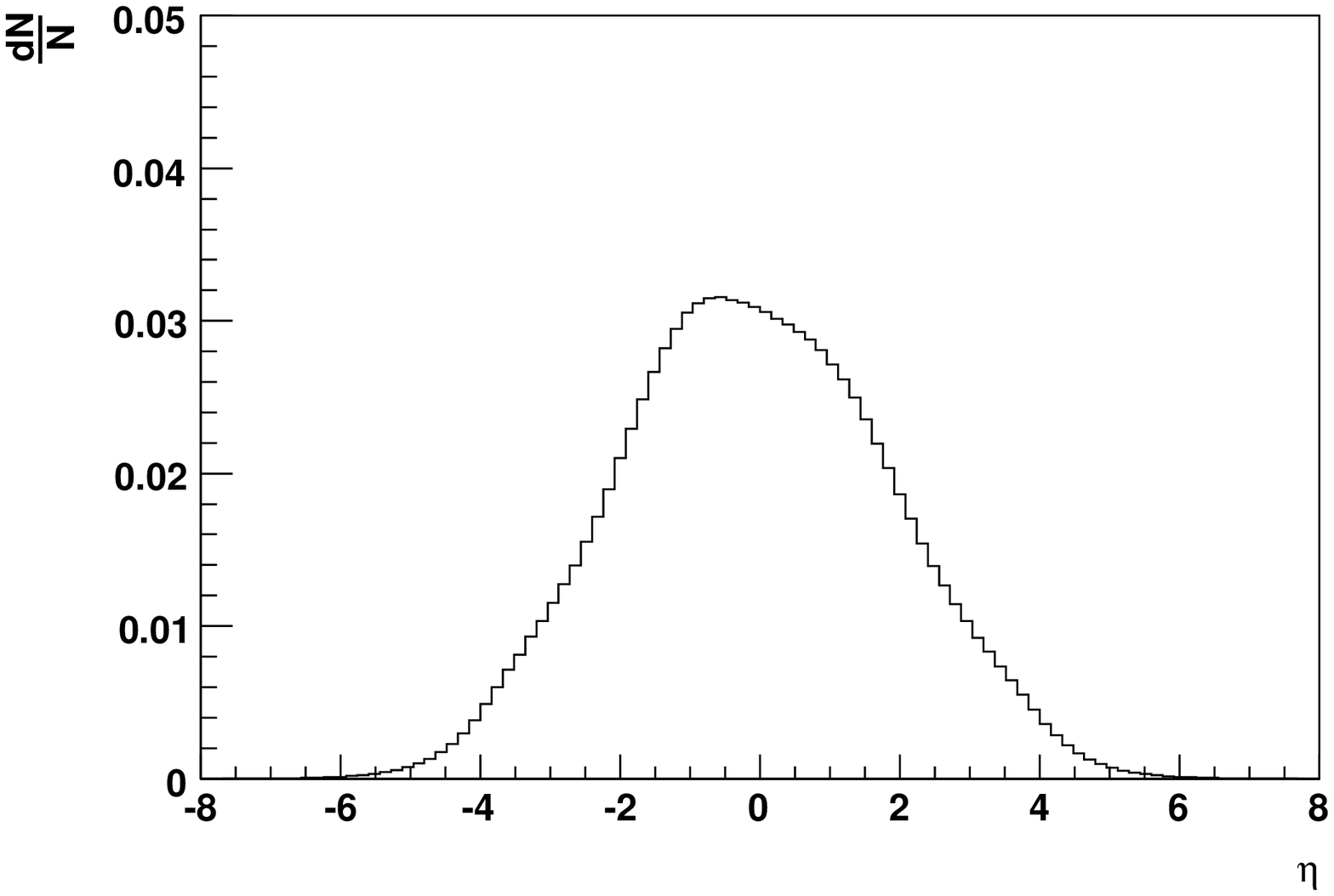}
\includegraphics[width=\picwidth]{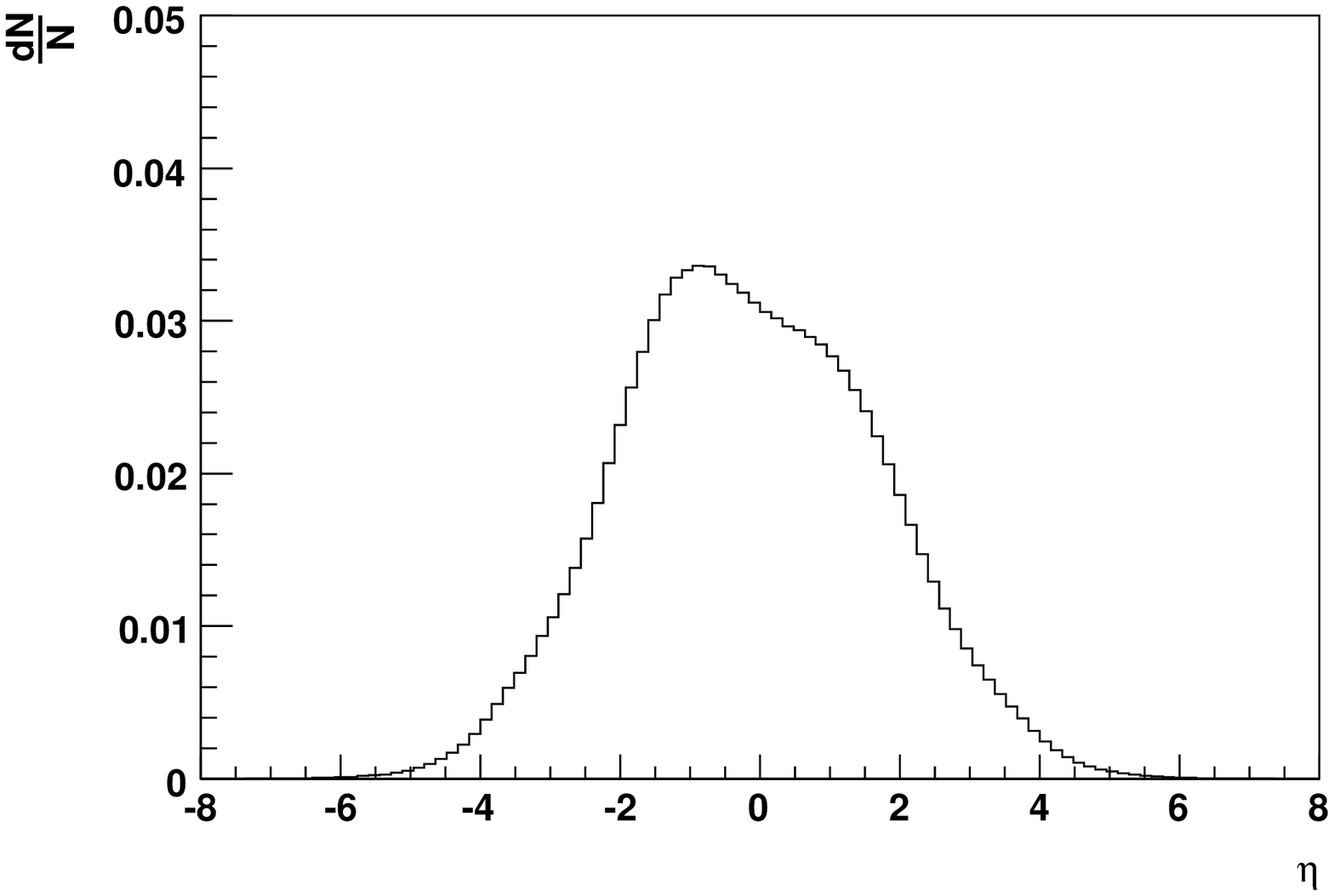}\\
\includegraphics[width=\picwidth]{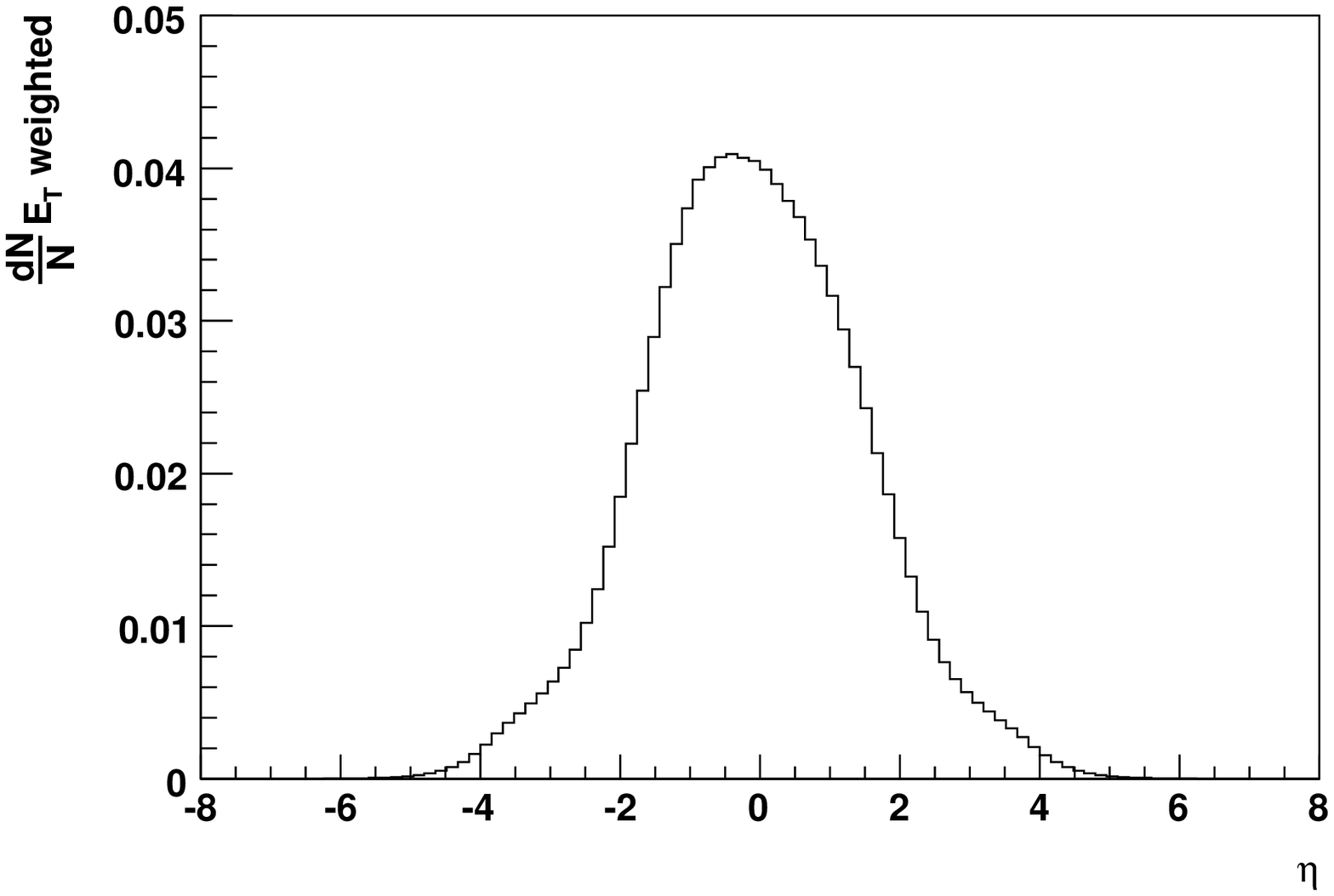}
\includegraphics[width=\picwidth]{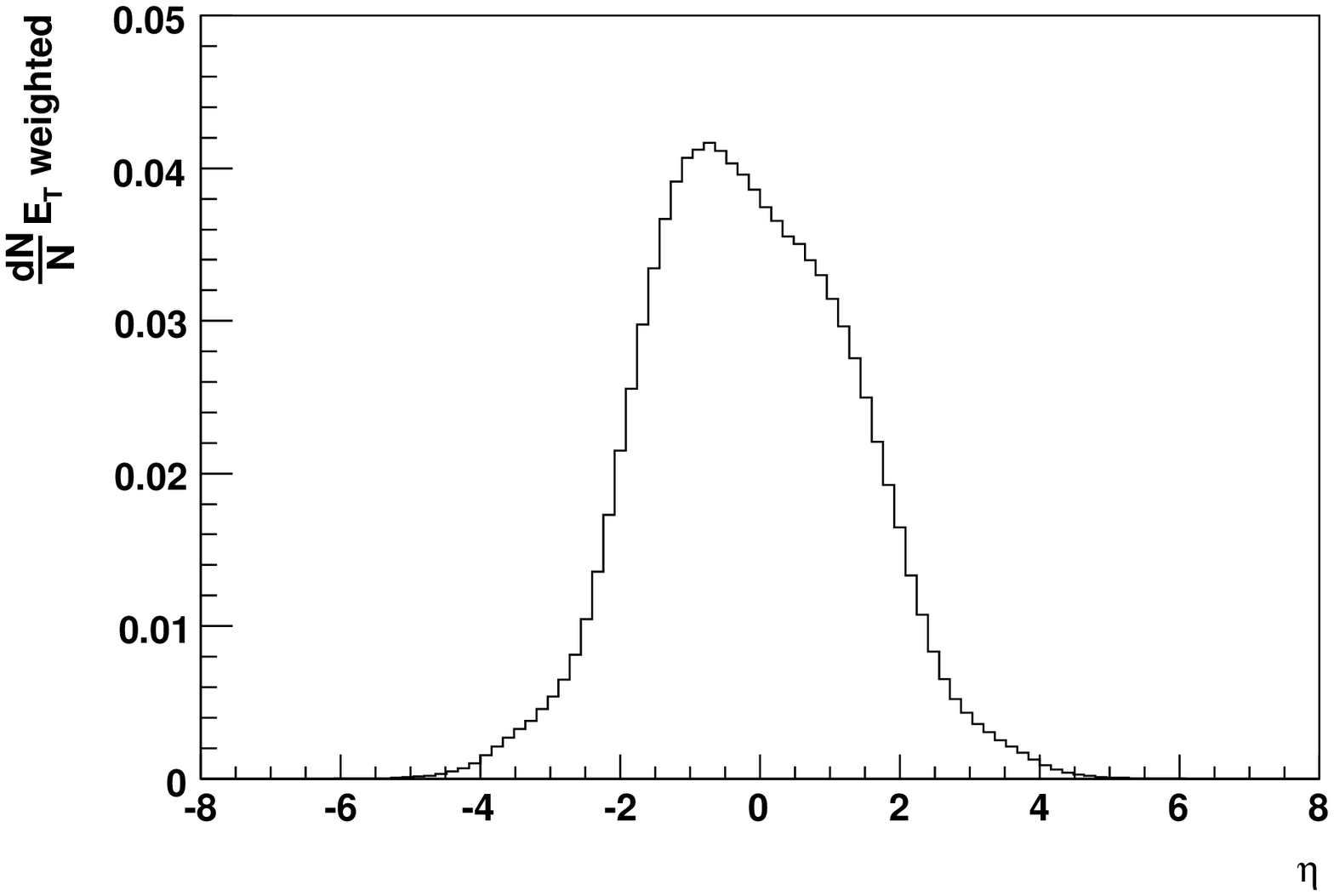}\\
\includegraphics[width=\picwidth]{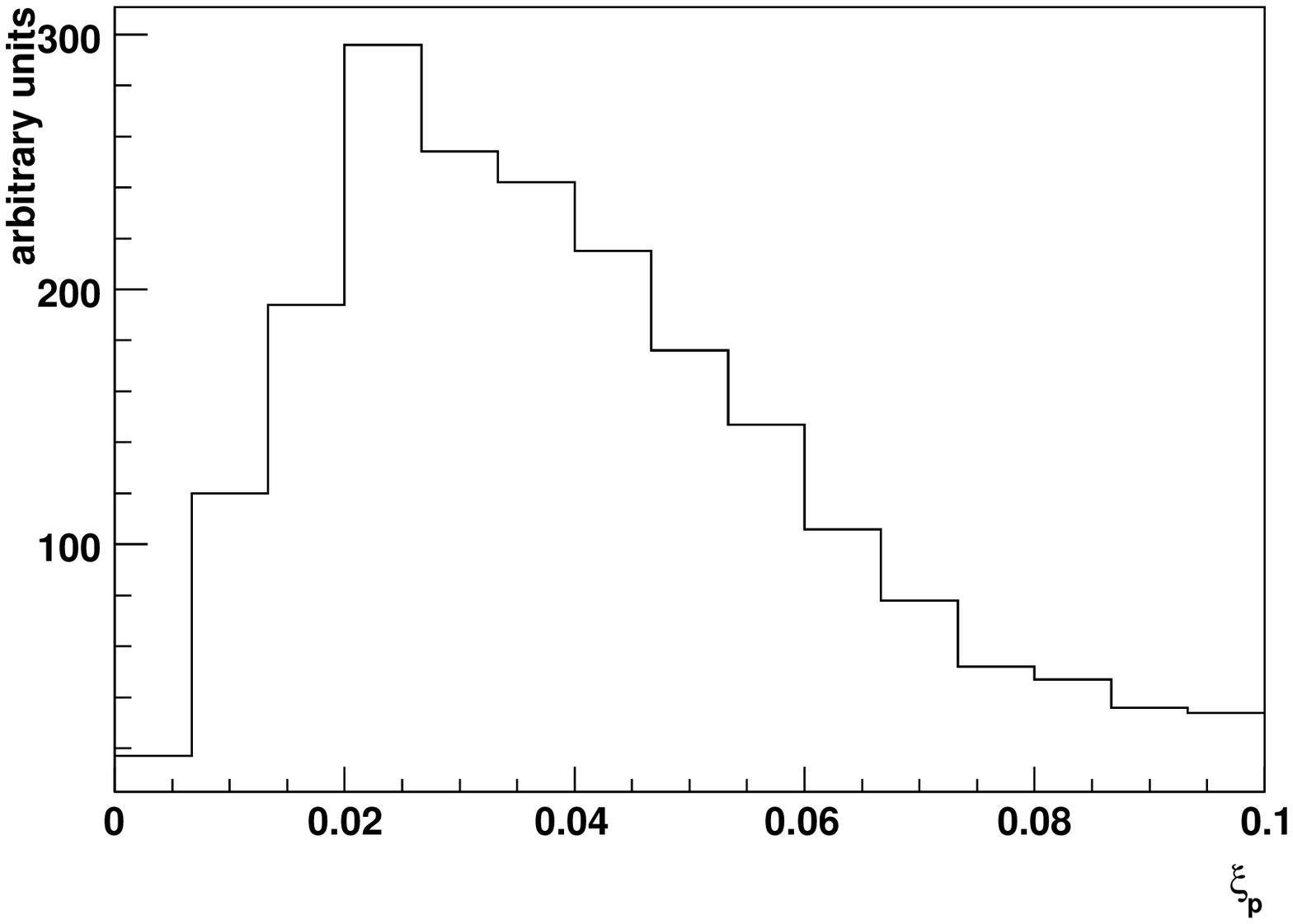}
\includegraphics[width=\picwidth]{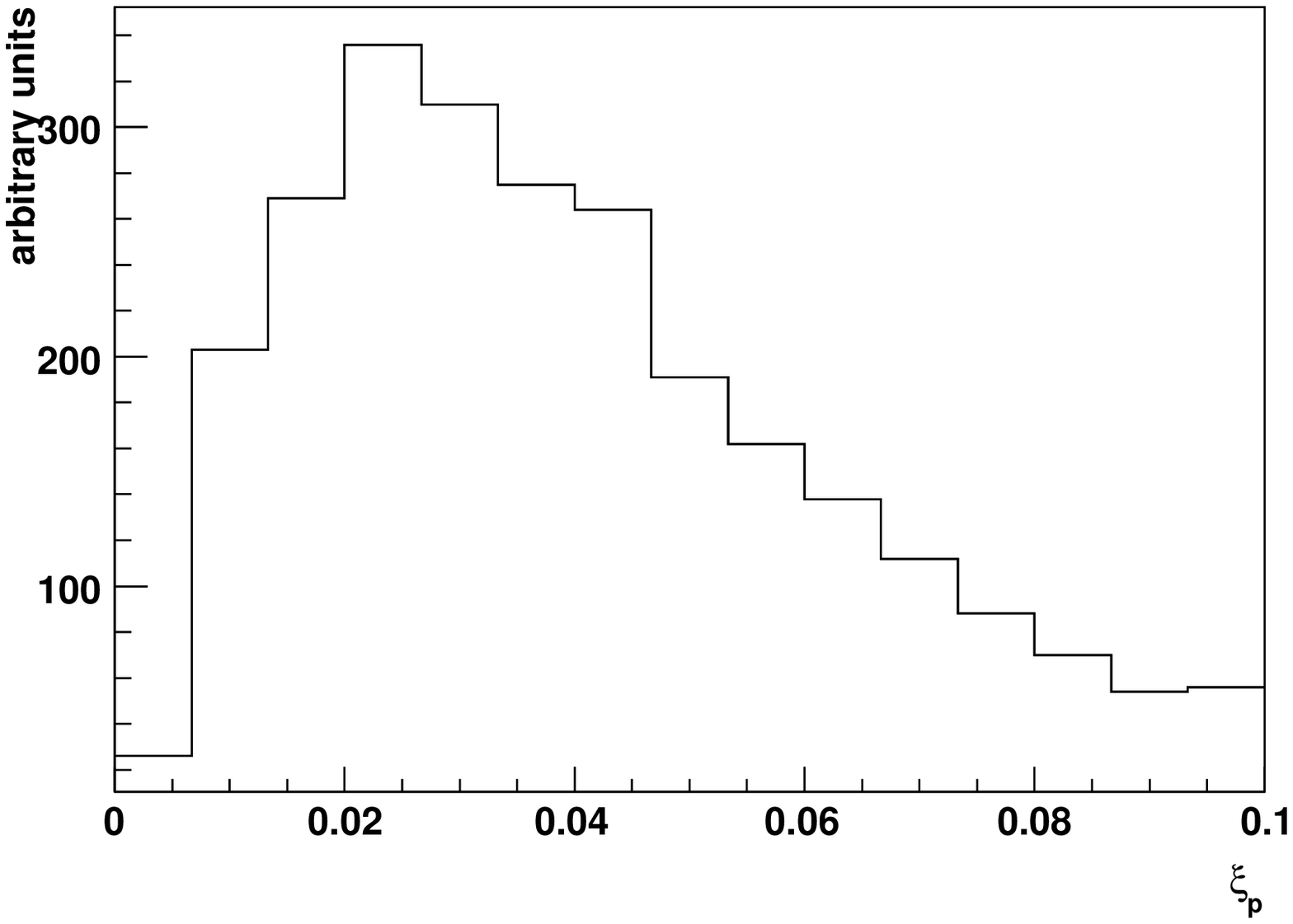}\\
\caption{Upper and medium plots: Rapidity and $E_T$ weighted rapidity 
distributions of all particles produced (except the protons); Lower plot:
momentum loss of the proton
in double pomeron exchange events $\xi_p$  for FM (left) and BL (right) 
inclusive models.}
\label{FigEnergy}
\end{figure}

\item A comparison between the proton momentum loss obtained from particles  $\xi_p^{part}$ calculated using
formula (\ref{eq:xipart}) and the proton momentum loss at generator level 
$\xi_p$ leads
to the factor 1.1 mentionned in a previous section. 
The dependance is displayed in Fig. \ref{Figetaxiinc}.
\begin{figure}
\includegraphics[width=\picwidth]{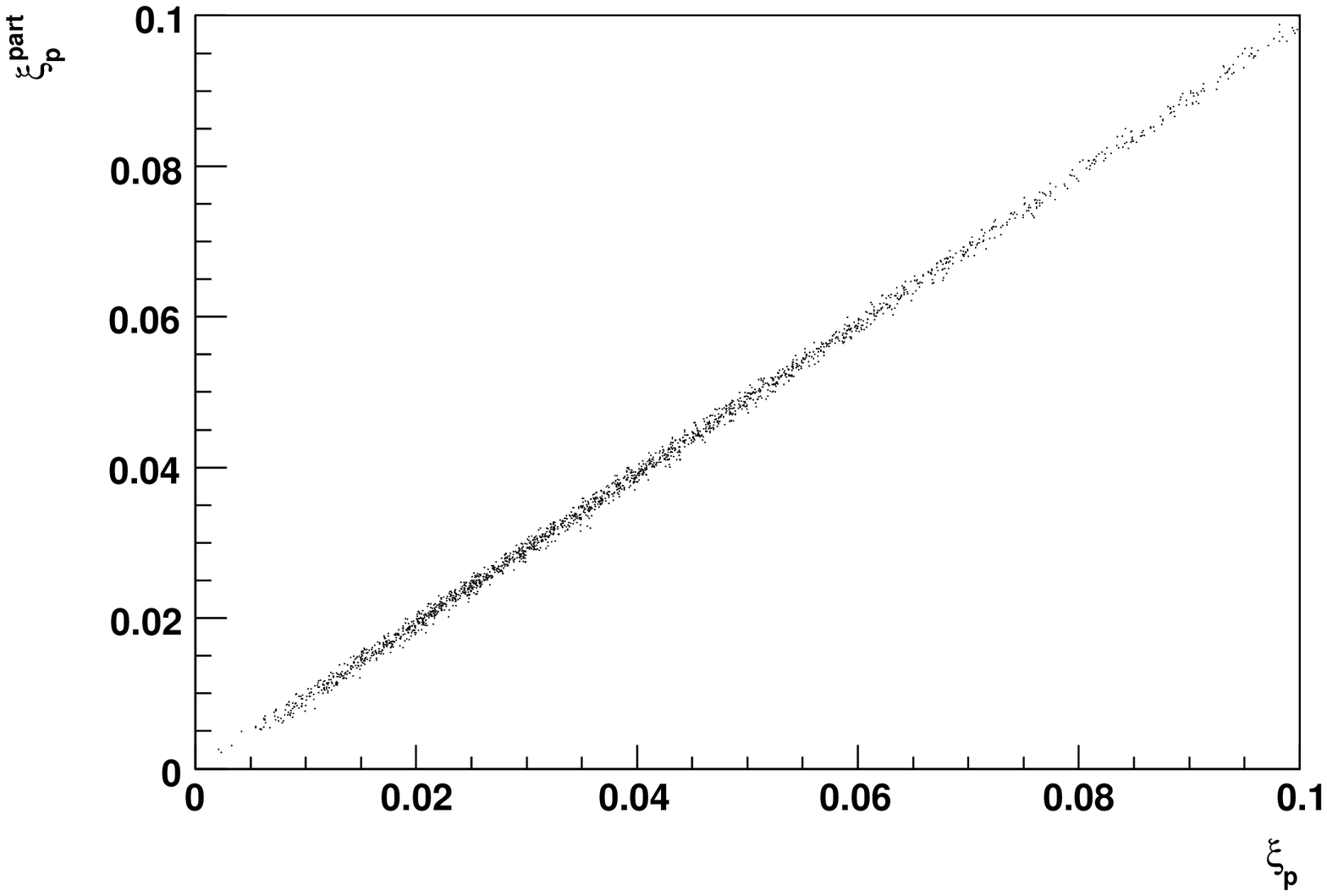}
\includegraphics[width=\picwidth]{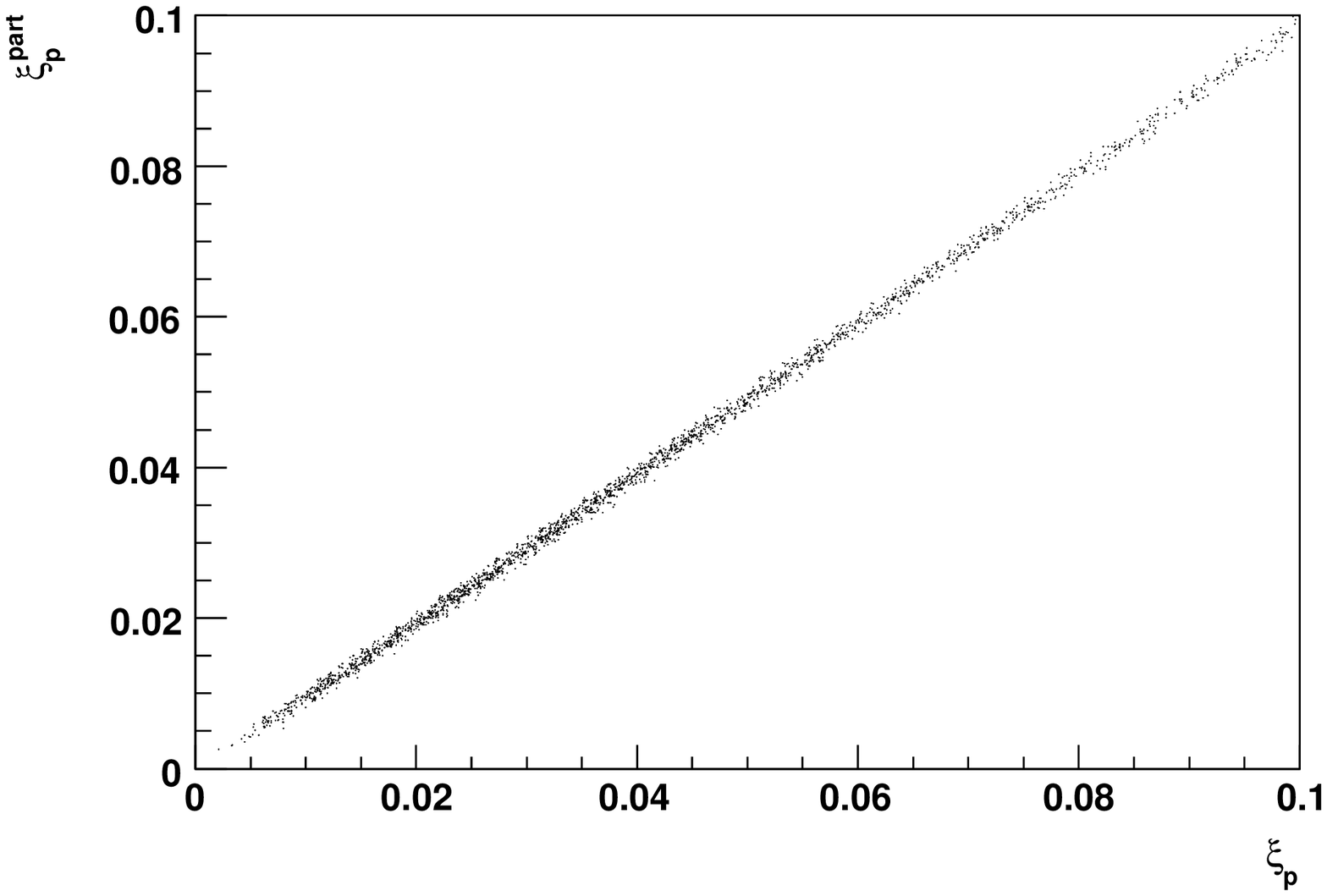}\\
\caption{Comparision of the proton momentum loss $\xi^{part}_p$ calculated 
with  formula (\ref{eq:xipart})
and the proton momentum loss $\xi_p$ at generator level.}
\label{Figetaxiinc}
\end{figure}

\item The size of the rapidity gap runs as a function of the momentum loss $\xi$ like 
$\Delta\eta\sim\log1/\xi$. The size of the gap which increases with 
decreasing $\xi$ for 
inclusive models can be seen in Fig. \ref{Figgapxi}. Regions of high rapidity
show the $\bar{p}$ hits whereas the low rapidity region is due to the 
produced particles 
detected in the central detector; they are well separated by a rapidity gap. 
For exclusive events, the size of a rapidity gap is larger and does not show
such a strong $\xi$
dependance as for inclusive models. 

\item The simulation interface plays a significant role in the determination of the exclusive contribution.
As previously stated, we cannot profit from having access to the full 
simulation interface
and having under control all the effects of the detector. In order to eliminate some effects
of the simulation we plot the dijet mass distribution $R_{JJ}$ using the information 
from the generator and check whether the need of exclusive events to describe the data
is still valid. Specifically, we require the same cuts as in Section 
\ref{sect:dmf} but 
the diffractive mass $M^{RP}$ is evaluted using the true (anti)protons momentum loss 
$(\xi_{\bar{p}})\xi_p$ at generator level  
\begin{equation}
M^{RP}_X=\sqrt{s\xi_{\bar{p}}\xi_{p}}.
\end{equation}
The dijet mass fraction calculated with $M^{RP}$ is shown in  Fig. \ref{FigDMFgen}. We see 
that the distribution is shifted to lower values of $R_{JJ}$, requesting 
slightly more exclusive 
events to describe the CDF data. The description of the data is 
also quite good.
\begin{figure}[h]
\includegraphics[width=\picwidth]{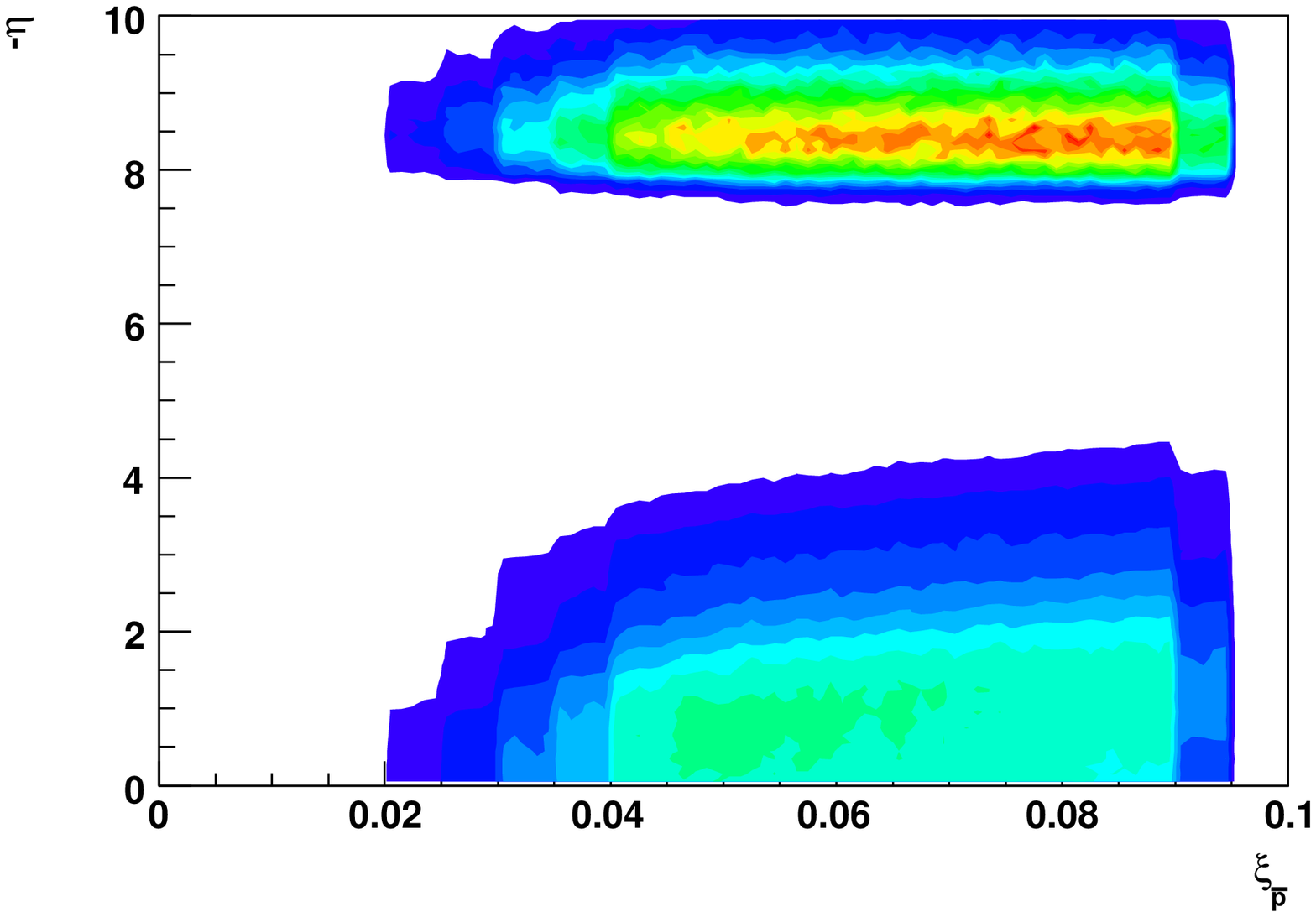}
\includegraphics[width=\picwidth]{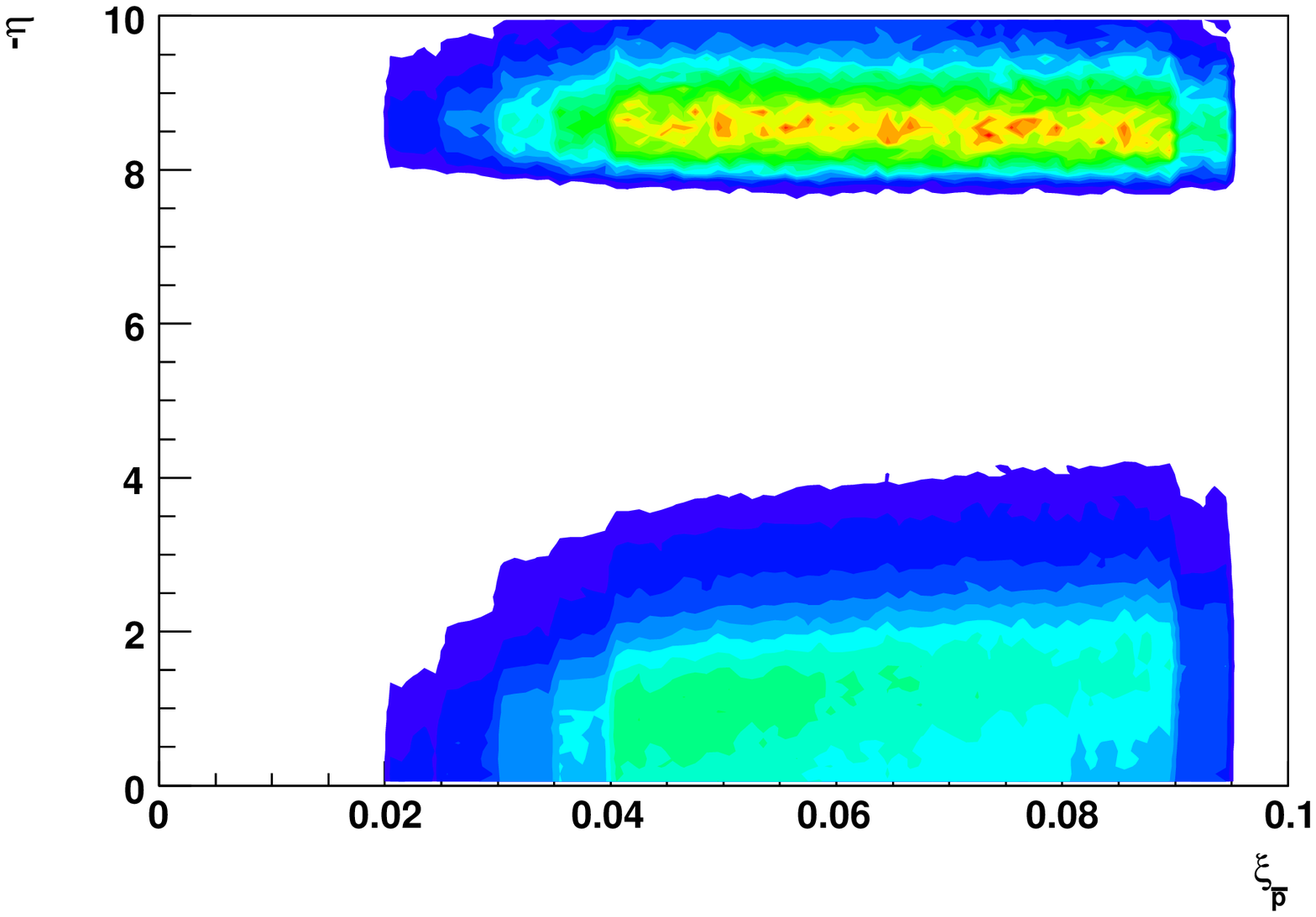}\\
\includegraphics[width=\picwidth]{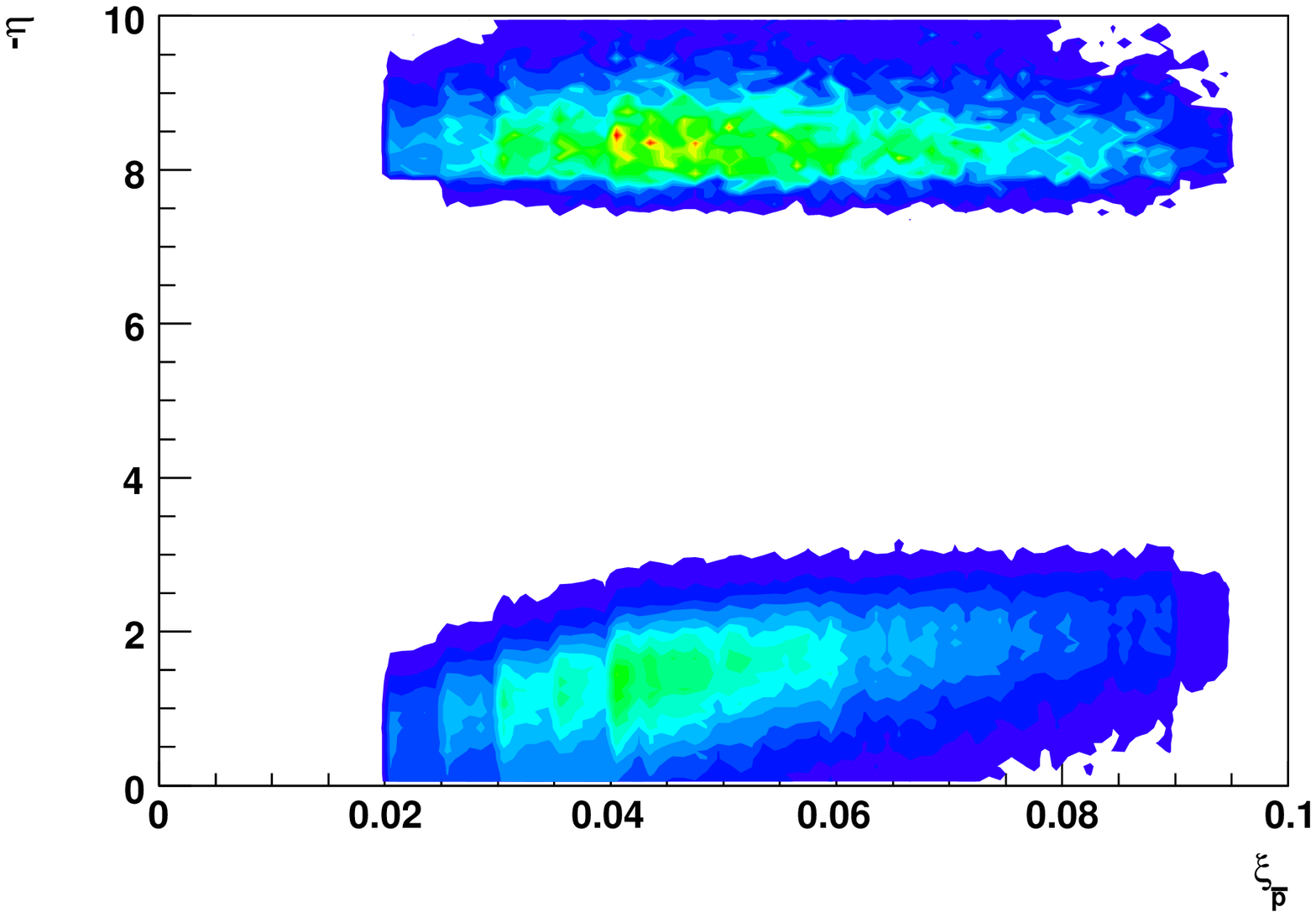}
\includegraphics[width=\picwidth]{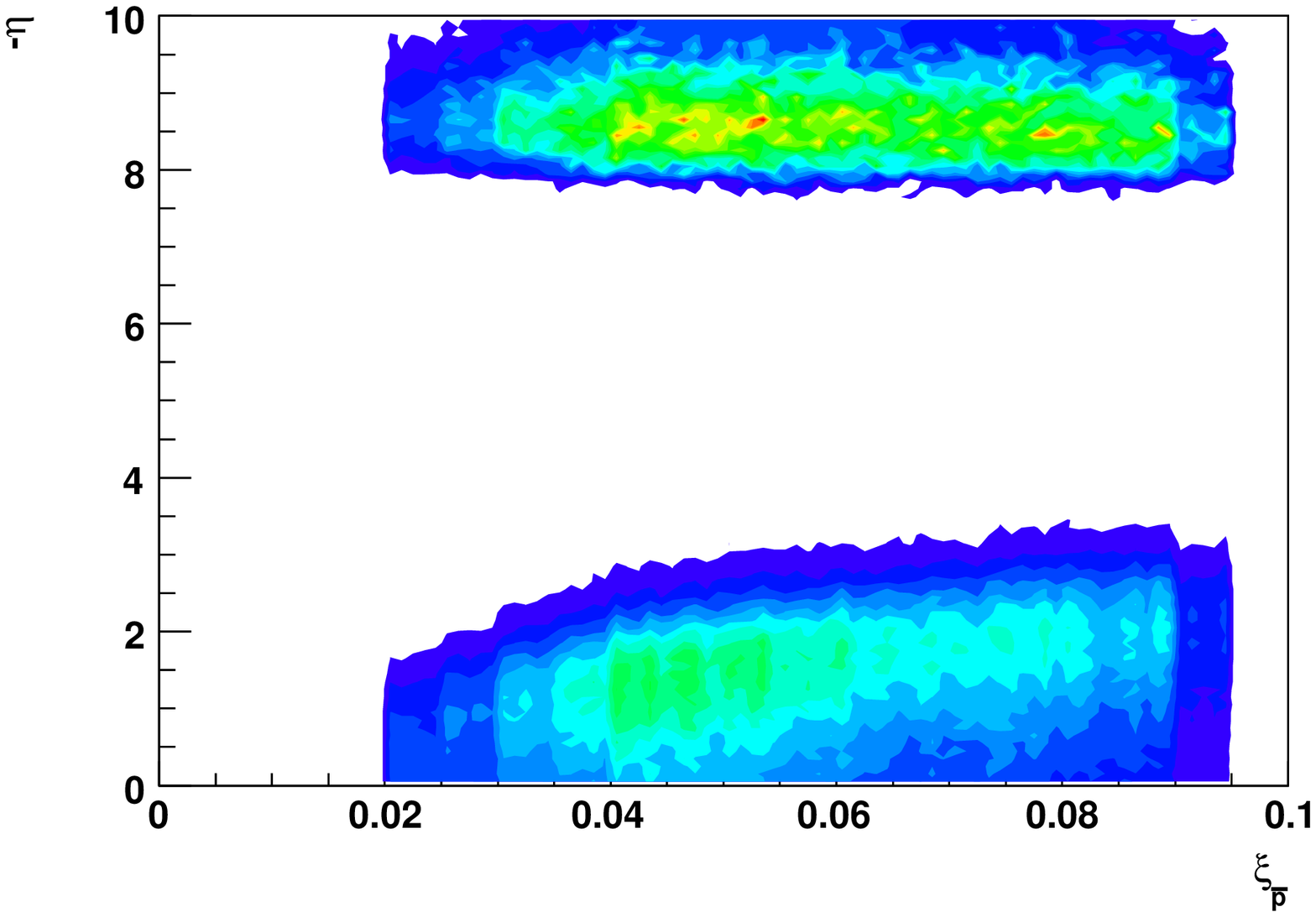}
\caption{Rapidity of particles on the $\bar{p}$ side vs. $\bar{p}$ momentum loss: inclusive models (top)
 for FM (left) and BL (right);
exlusive models (bottom) for KMR (left) and BL (right). Hits of scattered
$\bar{p}$ are included.}
\label{Figgapxi}
\end{figure}
\begin{figure}[h]
\includegraphics[width=\picwidth]{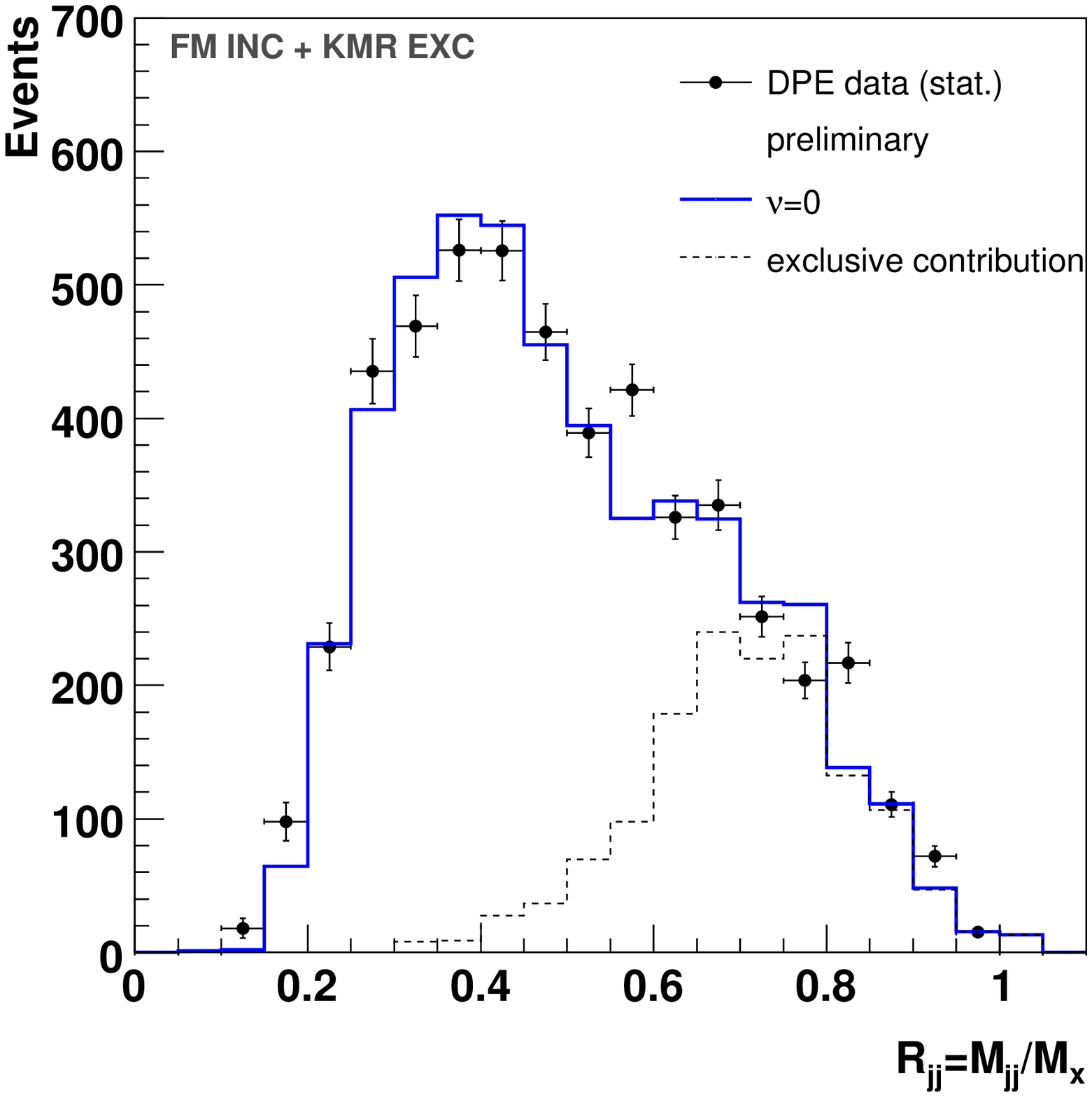}
\includegraphics[width=\picwidth]{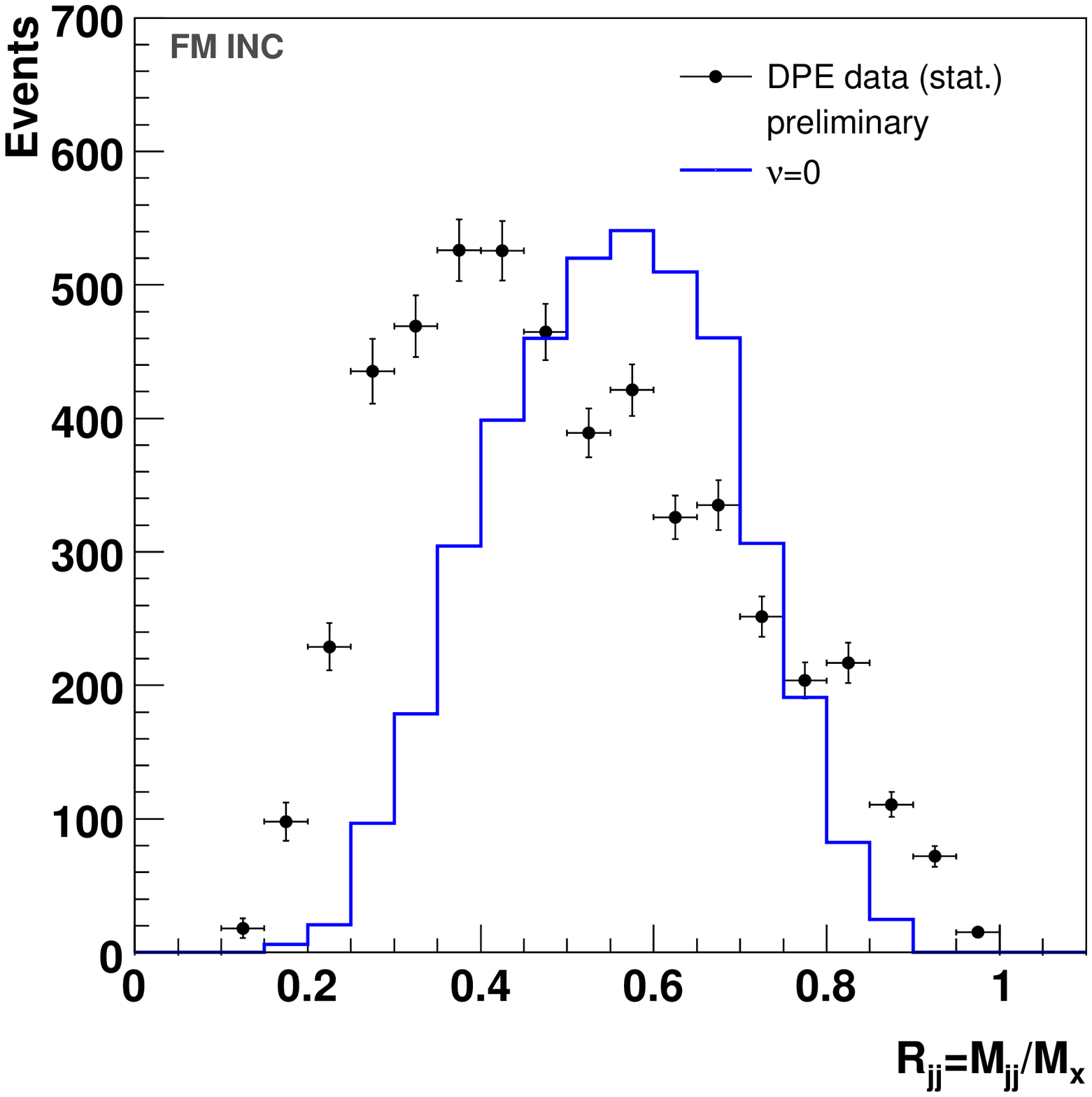}
\caption{Dijet mass fraction for jets $p_T>10\,\mathrm{GeV}$: FM + KMR (left), 
and at generator
level calculated according to (\ref{eq:dmfbeta}). }
\label{FigDMFgen}
\end{figure}

\item The role of the simulation interface to reconstruct jets can be illustrated by comparing the 
above distributins to DMF calculated at generator level defined as
\begin{equation}
R_{JJ}=\frac{M_{JJ}}{M_X}=\frac{\sqrt{s\xi_{\bar{p}}\xi_p\beta_1\beta_2}}{\sqrt{s\xi_{\bar{p}}\xi_p}}=\sqrt{\beta_1\beta_2},
\label{eq:dmfbeta}
\end{equation}
where $\beta_1$, $\beta_2$ denote the fraction of the pomeron carried by the interacting parton. As can be seen, 
in Fig. \ref{FigDMFgen} (right), the DMF distribution at pure generator level shows a completely different shape 
not compatible with CDF data and shows the importance of jet reconstruction.  

\end{itemize}

\end{document}